\pdfoutput=1

\documentclass[final,3p,times,twocolumn]{elsarticle}
\usepackage{lineno}

\usepackage{amsmath}

\journal{Nucl. Instr. Meth. Phys. Res.}
\begin{document}
\begin{frontmatter}


\author[ks]{S.~Kunwar\corref{cor1}}
\cortext[cor1]{Corresponding Author.}
\ead{samridhak@gmail.com}
\author[ut]{R.~Abbasi}
\author[ks]{C.~Allen}
\author[ut]{J.~Belz}
\author[ks,mep]{D.~Besson}
\author[ut]{M.~Byrne}
\author[ut]{B.~Farhang-Boroujeny}
\author[bg]{W.H.~Gillman}
\author[ut]{W.~Hanlon}
\author[ks]{J.~Hanson}
\author[ut]{I.~Myers}
\author[mep]{A.~Novikov}
\author[ks]{S.~Prohira}
\author[ks]{K.~Ratzlaff}
\author[ut]{A.~Rezazadeh}
\author[ut]{V.~Sanivarapu}
\author[ut]{D.~Schurig}
\author[mep]{A.~Shustov}
\author[mep]{M.~Smirnova}
\author[bh]{H.~Takai}
\author[ut]{G.B.~Thomson}
\author[ks]{R.~Young}
\address[ut]{University of Utah, Salt Lake City, UT 84112 U.S.A.}
\address[ks]{University of Kansas, Lawrence, KS 66045 U.S.A.}
\address[bg]{Gillman \& Associates, Salt Lake City, UT 84106 U.S.A.}
\address[us]{Utah State University, Logan, Utah 84322 U.S.A.}
\address[bh]{Brookhaven National Laboratory, Upton, NY 11973 U.S.A.}
\address[mep]{National Research Nuclear University MEPhI (Moscow Engineering Physics Institute), Moscow 115409 Russia}

\title{Design, Construction and Operation of a Low-Power, Autonomous Radio-Frequency Data-Acquisition Station for the TARA Experiment}

\begin{abstract}
Employing a 40-kW, 54.1 MHz radio-frequency transmitter just west of Delta, UT, the TARA (Telescope Array RAdar) experiment seeks radar detection of extensive air showers (EAS) initiated by ultra-high energy cosmic rays (UHECR). For UHECR with energies in excess of $10^{19}$ eV, the Doppler-shifted ``chirps'' resulting from EAS shower core radar reflections should be observable above background (dominantly galactic) at distances of tens of km from the TARA transmitter. In order to stereoscopically reconstruct cosmic ray chirps, two remote, autonomous self-powered receiver stations have been deployed. Each remote station (RS) combines both low power consumption as well as low cost. Triggering logic, the powering and communication systems, and some specific details of hardware components are discussed.
\end{abstract}


\end{frontmatter}


\section{Introduction}
Within the last 30 years, the sub-field of ultra-high energy cosmic ray (UHECR; $E>10^{18}$ eV) astronomy has emerged as a vibrant experimental and theoretical sub-field within the larger field of particle astrophysics, comprising studies of both charged and neutral particles at macroscopic kinetic energies.
The physics interest in UHECR lies in understanding i) the nature of the cosmic accelerators capable of producing such enormously energetic particles at energies millions to billions of times higher than we are capable of producing in our terrestrial accelerators, ii) the details of the interaction of UHECR with the cosmic ray background, evident in the observed energy spectrum of cosmic rays as an upper `cut-off'\cite{GZK}, or maximum observed energy, at approximately $10^{20}$ eV, and iii) correlations in the arrival directions of UHECR with exotic objects such as neutron stars, gamma-ray bursts (GRB), and active galactic nuclei (AGN). Experimentally, charged-particle UHECR astronomy
is currently dominated by two experiments -- the Southern Hemisphere Pierre Auger Observatory (PAO)\cite{PAO} based in Malargue, Argentina and the Northern Hemisphere Telescope Array (TA)\cite{TA} based in Utah, USA. The construction of these observatories is very similar, based on a large number of ground detectors sampling the charged component of Extensive Air Showers (EAS) and deployed over hundreds of square kilometers at $\sim$1 km spacing, coupled with a much smaller number of atmospheric nitrogen fluorescence detectors having a much more restricted duty cycle, but individually capable of providing a more comprehensive image of atmospheric shower development. Within the last decade, the PAO has been complemented by an array of radio-wave antennas, capable of measuring the signal generated primarily by the separation of the charged particles comprising the down-moving air shower in the geomagnetic field\cite{AERA}.

\begin{figure}
\centerline{\mbox{\includegraphics[width=0.48\textwidth]{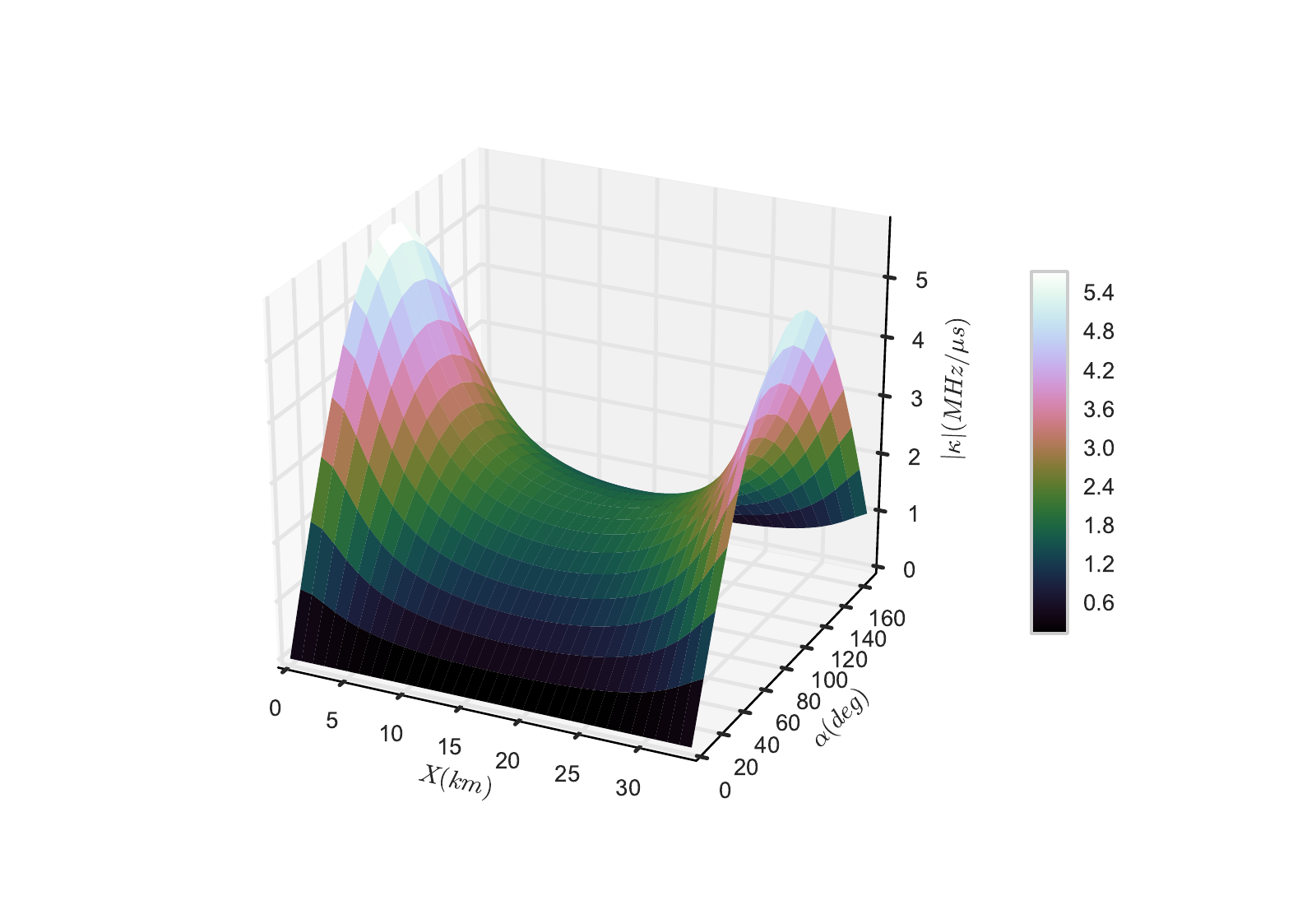}}}
\caption{Dependence of chirp rate on the distance from the receiver to the shower core at shower max
and also the angle between the shower direction and a line joining the receiver to shower max, from geometric considerations only.
\label{fig:chirprate}}
\end{figure}

The TARA experiment (Telescope Array RAdar) attempts to achieve large aperture for EAS detection by bi-static observation of radar reflections from the core of an EAS at radio frequencies. The radial charge density of a developing EAS varies roughly with distance; within a radial distance of roughly 1 cm of the shower axis, the charge density within that shower core is sufficiently high that a radio-reflective plasma can develop. If that plasma is sufficiently long lived, it will coherently scatter radio signals in the forward direction; such forward scattered radio signals are, in principle, detectable at distances of tens of km. Straightforward geometric considerations lead to the expectation that, since the total pathlength from 54.1 MHz transmitter$\to$shower core$\to$receiver typically decreases with time for a down-moving shower, the received radar echo Doppler-redshifts by approximately 50\% over a period of several microseconds, resulting in a radio-frequency ``chirp'' with slope of order several MHz per microsecond. The chirp rate, in terms of the distance from shower to receiver, and also the relative inclination angle of the shower is a strong function of geometry, as illustrated in 
Fig.~\ref{fig:chirprate}.

\begin{figure}
\centerline{\mbox{\includegraphics[width=0.48\textwidth]{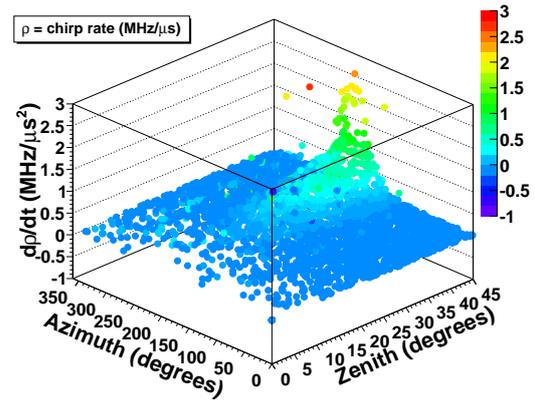}}}
\caption{Dependence of time-derivative of chirp rate (shown in both color scale and vertical axis) as a function
of incident cosmic ray geometry for a large number of Monte Carlo simulated radar
echoes. In the Figure, the azimuthal angle $\phi$=0 is defined by the line between
the transmitter and receiver; $\theta$ is the conventional shower zenith angle.
\label{fig:crate}}
\end{figure}

The chirp can be characterized by several parameters, including the average chirp slope [MHz/s], the rate of change in the chirp slope (i.e, the time derivative of the slope [MHz/s/s]), as well as the initial, final, and maximum detectable chirp rate values. As illustrated in Figure \ref{fig:crate}, the change in the chirp slope is also 
tightly coupled to event geometry, as expected, so that the event geometry can, to some extent, be constrained by the observed chirp parameters.

\subsection{Paper Outline}
This paper is structured as follows. We first outline the general features of the expected signal characteristics, which form the basis for our Monte Carlo simulations. We then review the evolution of the TARA project, with particular attention given to the remote station concept and hardware details. The performance of the current two-station configuration, including a local calibration source is then described, concluding with a prospectus for the future.

\section{Chirp Simulation}
To characterize the expected signals, the
TARA remote station data is simulated \cite{Helio,Isaac14} by accounting for the effects of the shower properties and event geometry, atmospheric properties including local gas density, and the characteristics of the receiving antenna. Once the primary energy and the geometry are specified, the shower properties (such as shower maximum - \emph{$X_{max}$}) are derived from the Gaiser-Hillas prescription\cite{GH}.  The radio cross-section of the shower (RCS) is affected by the event geometry, and is modeled as a thin wire with a radius of order 1--2 cm.  The length of the wire is governed by the plasma charge density\cite{Gorham}.  The induced plasma charge density and lifetime, in turn, are determined by several effects, the strongest of which are collisional damping of the plasma electrons with atmospheric molecules, and attachment of the plasma electrons to oxygen molecules through two and three-body attachment mechanisms\cite{Isaac14,SamThesis}.  Accounting for these effects at typical air shower altitudes yields plasma lifetimes of 10-100 ns.

The radiation patterns of the receiving and transmitting antenna arrays are incorporated into the simulation after careful field and anechoic chamber measurements of both the angular orientation and frequency dependence\cite{Isaac14}.  The phase and amplitude of the transmitted 54.1 MHz carrier are tracked, and updated once the observable RCS of the air shower enters the Fresnel zone of the transmitter-receiver system. The frequency evolution of the reflected signal is also directly tracked, based on the non-relativistic change in path length from transmitter to shower core to receiver.

\subsection{Performance Estimate from Simulations}
To derive event parameters such as zenith angle, a neural network analysis has been built around the output waveform properties.
The power of the reflected signal versus time is governed by the development of the RCS and the bi-static radar equation, which is then modulated by the phase to produce the final waveform in the receiver.  The simulation predicts 100-1000 cm$^{2}$ for the RCS, depending on parameters like the zenith angle, and the carrier incident angle with respect to the thin wire. To provide a realistic sampling of potentially detectable air showers, three thousand Telescope Array surface detector (SD) events were passed as inputs to the chirp simulation, which requires zenith angle, azimuthal angle, and energy. 

\begin{figure}
\centering
\includegraphics[width=0.48\textwidth]{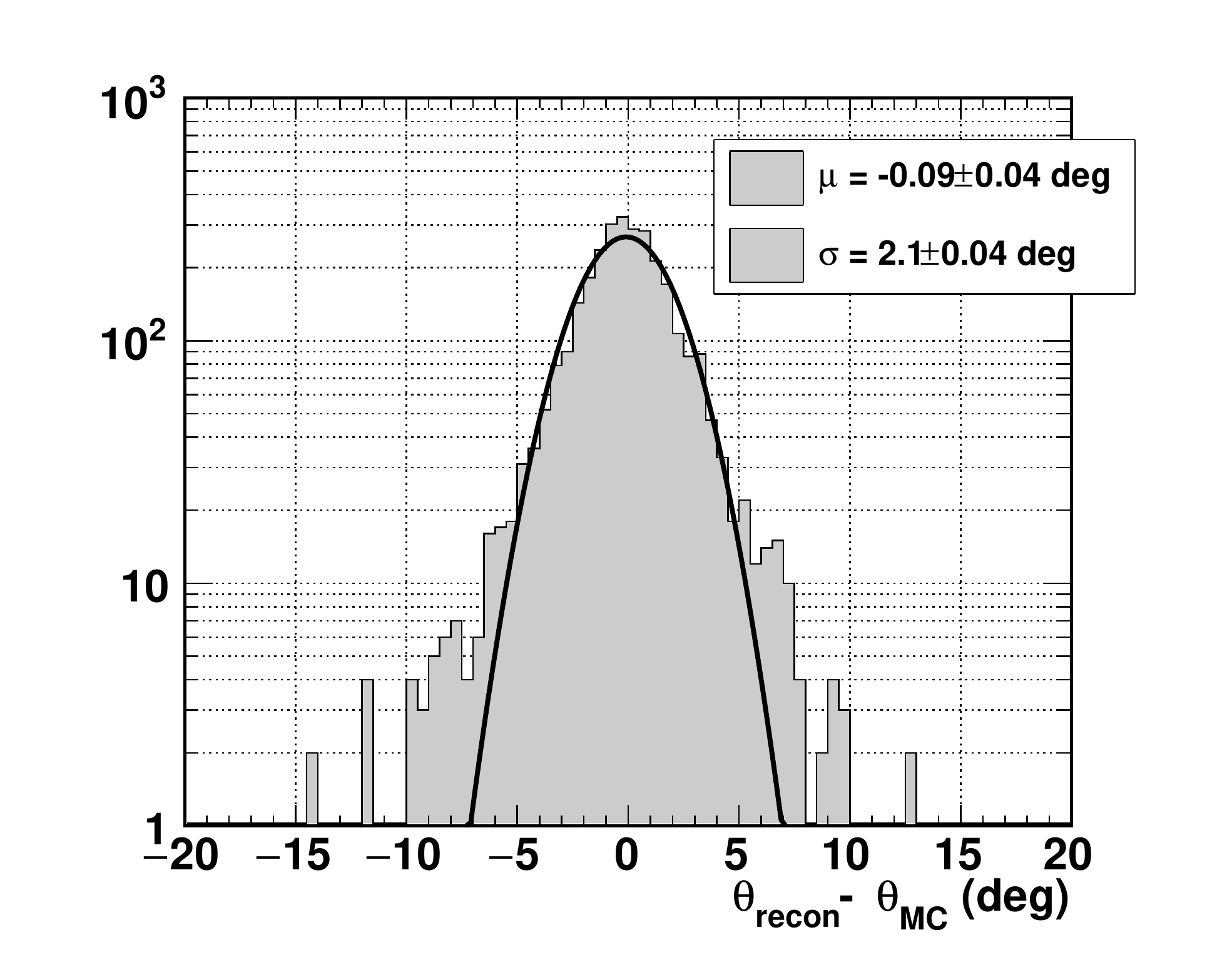}
\caption{Difference between MC truth zenith angle, $\theta_{MC}$, and the reconstructed zenith angle $\theta_{recon}$, after the neural network is trained.  There are 80 0.5 degree width bins in the plot.  \label{jchFig}}
\end{figure}

The simulated chirp waveforms are then de-chirped as in the remote stations (delaying a copy and mixing to find low-frequency monotone, as detailed below).  The chirp rate, maximum voltage, most powerful frequency, and chirp rate derivative, as measured by a single simulated station with vertical and horizontally polarized receivers, were tracked versus time for each event.  Those parameters are passed finally to a 30-node by 4-layer neural network, with, e.g., in the case of Fig \ref{jchFig}, the target parameter being the shower zenith angle.  The neural network uses the Levenberg-Marquardt training technique, and attempts to minimize the mean-squared error between Monte Carlo truth and network output, using between 40-80 percent of the data to train.  The final error (deviation between input and reconstructed zenith angle) histogram is shown in Fig.~\ref{jchFig}, and the width of this histogram varies from $2^{\circ}-3^{\circ}$, when the percentage of the data used for training is adjusted from 40-80 percent, respectively.  The Gaussian fit to the results indicate little systematic bias, and that degree-level precision is available with one station, reading out both Vertical as well as Horizontal polarizations.  

\section{TARA Development Timeline}
Initiated in 2011 with a 2 kW transmitter, TARA upgraded to a 40-kW transmitter in 2013, operating at a carrier frequency of 54.1 MHz. Within the last two years, concerted studies of background and calibration of performance have been conducted and documented\cite{Isaac14}, using the primary receiver system located at the site of the Telescope Array fluorescence detectors. 

To enhance the sensitivity of TARA, at a location largely free of episodic impulsive backgrounds, we have designed and deployed two solar-powered, autonomous remote stations (RS), each operating with a power draw of approximately 15 Watts. 
Each remote receiver station is designed to detect chirp echoes with signal to noise ratios (SNR) as low as -10 dB. 

\section{Remote Station Design}
The remote stations combine a custom front-end antenna with a custom data acquisition system, with separate logic for event triggering and waveform capture. Each station also includes both a power source as well as the capability to broadcast captured data back to the primary Telescope Array fluorescence detector site via a GHz communications data link.

\begin{figure}
\centerline{\mbox{\includegraphics[width=0.48\textwidth]{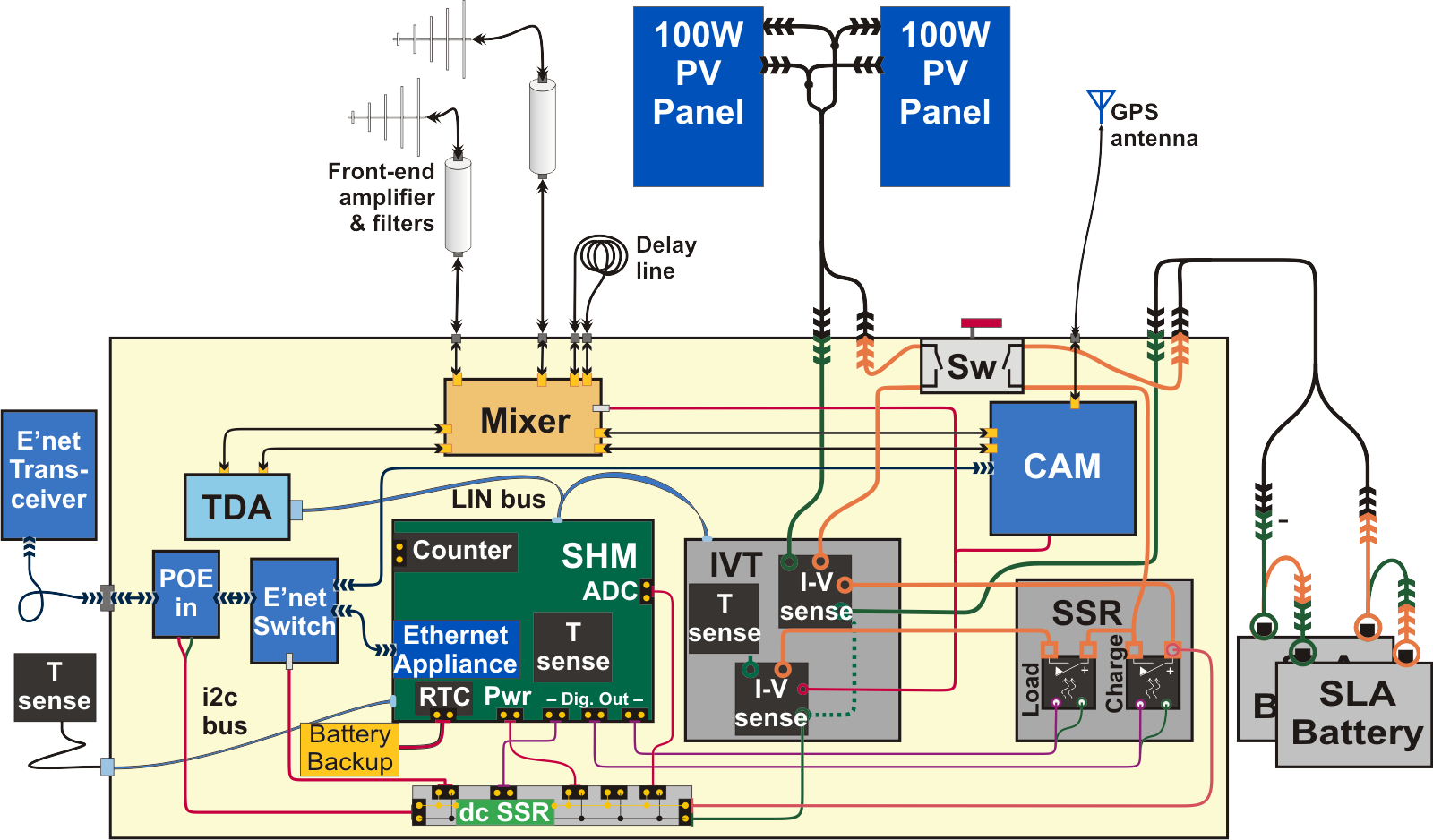}}}
\caption{Block diagram showing the different components of the TARA RS detectors. 
\label{fig:TARA_Block_dia}}
\end{figure}

The first element of each RS is the front-end Log Periodic Dipole Antenna (LPDA), with outputs separately for VPol and also HPol signal. Beyond this, each RS comprises five basic components: the Mixer Module to translate an input chirp signal into a triggerable monotone, the Chirp Acquisition Module (CAM) to record input chirp waveforms having sufficiently high Signal-to-Noise Ratio (SNR), a Transient Detector Apparatus (TDA, which simply registers the number of excursions over a ten-second interval above a pre-set voltage threshold) to provide simplified threshold-crossing information for the input signals, and two slow-control devices, including the IVT (Current Voltage Temperature) Board and the overall System Health Monitor (SHM) electronics. An RS block diagram is presented in Fig.~\ref{fig:TARA_Block_dia}. 

\begin{figure}
\centerline{\mbox{\includegraphics[width=0.48\textwidth]{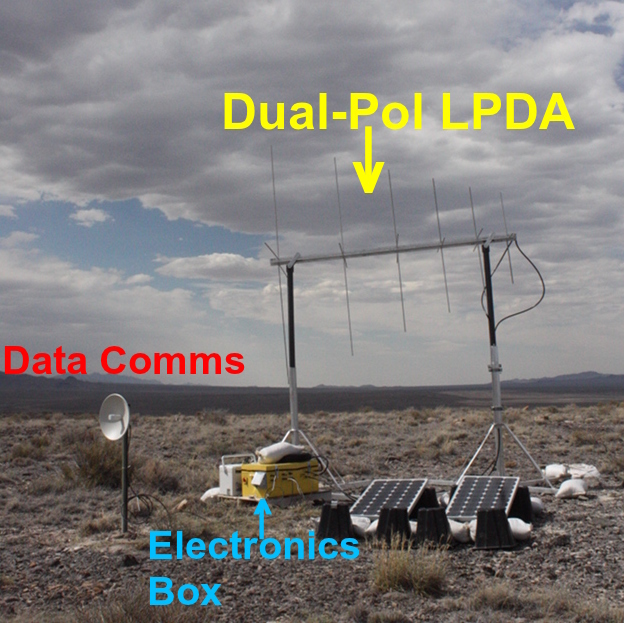}}}
\caption{Remote Station-1 (RS-1), prior to displacing the electronics box from the LPDA.
\label{fig:RS1}}
\end{figure}

\begin{figure}
\centerline{\mbox{\includegraphics[width=0.48\textwidth]{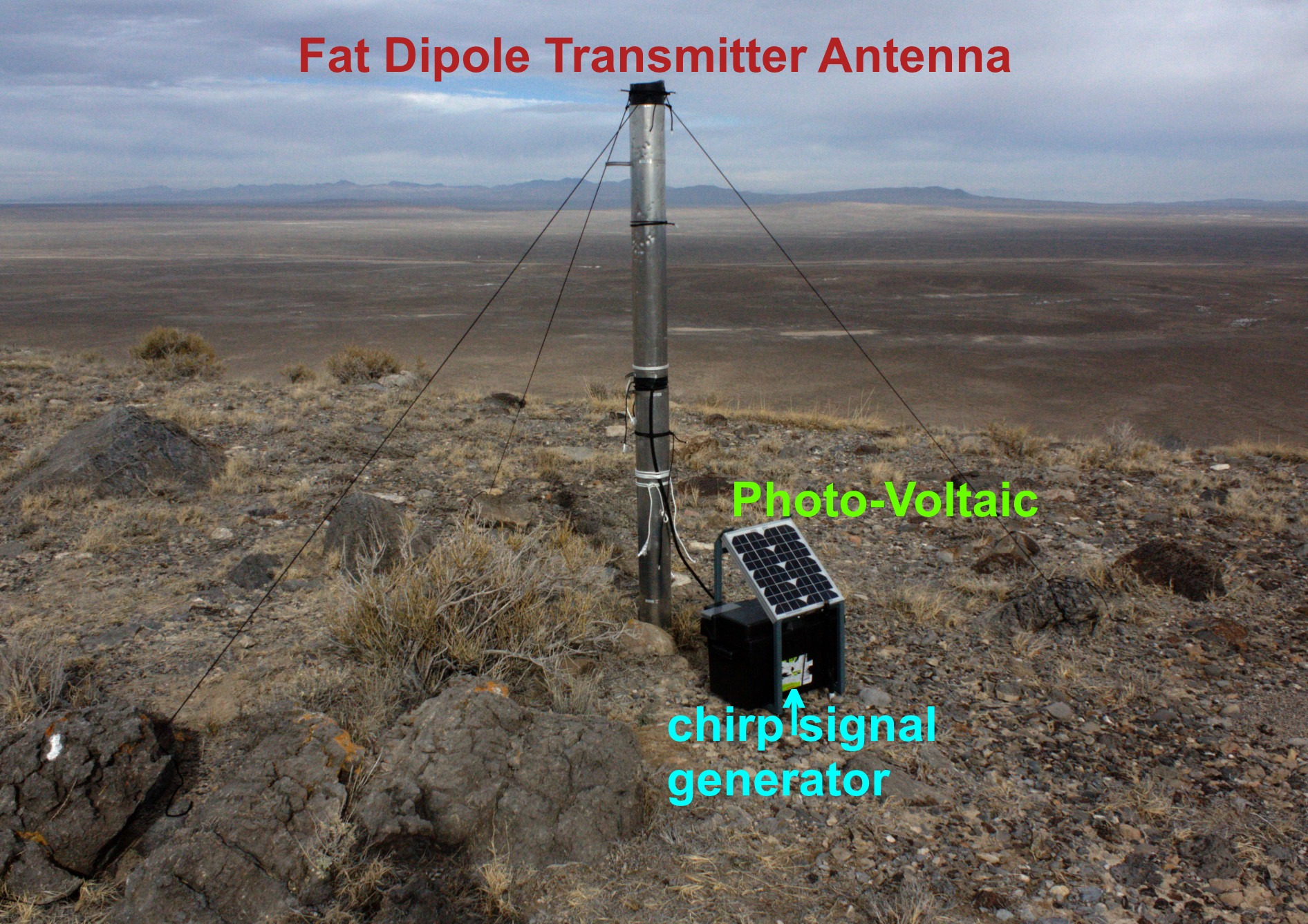}}}
\caption{Chirp Calibration Unit (CCU), showing the fat dipole used as a transmitter antenna, and fixed in place with guy wires, plus the photovoltaic (PV) panel
powering the chirp signal generator (in box below PV).
\label{fig:CCU}}
\end{figure}

\begin{figure*}
\centerline{\mbox{\includegraphics[width=0.98\textwidth]{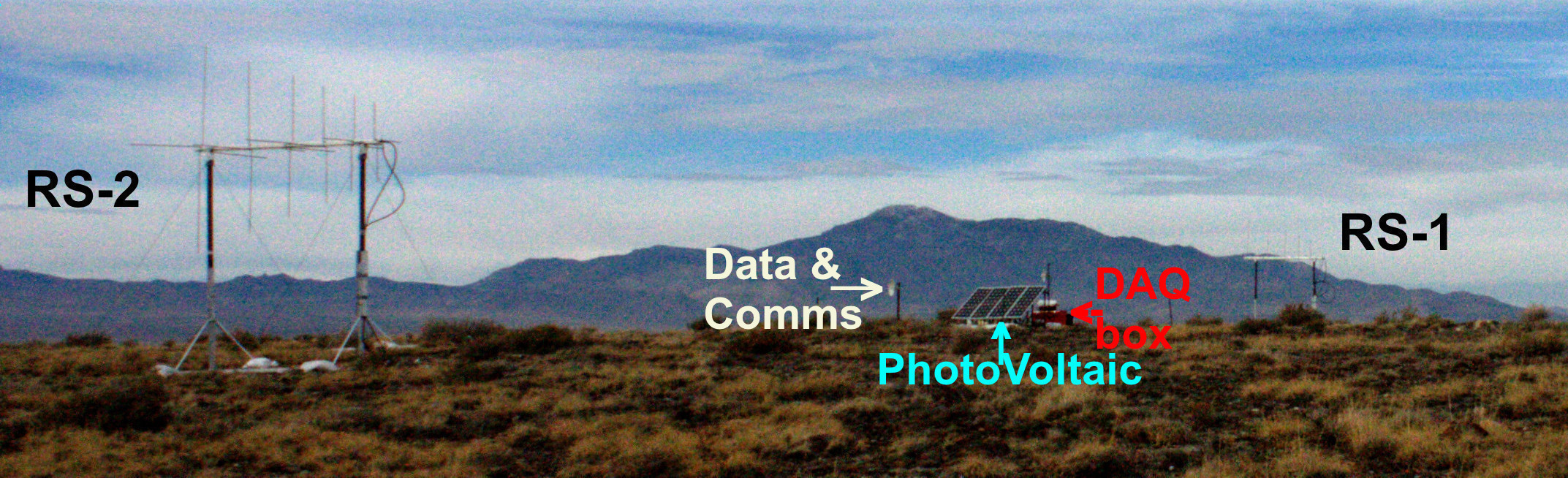}}}
\caption{RS-1 (right) and RS-2 (left).
\label{fig:RS1_RS2}}
\end{figure*}

Development of the RS system was completed in three stages: i) installation of RS-1 on June 28, 2014 at latitude=39.2427 N, longitude=113.0896 W, approximately 5 km from the main Telescope Array fluorescence detector site (Fig.~\ref{fig:RS1}), 
ii) installation of RS-2 on November 28, 2014 approximately 60 m displaced to the southeast of RS-1, and iii) installation of a local Chirp Calibration Unit (CCU; Fig.~\ref{fig:CCU}) to provide a continuous calibration source, on January 19, 2015, located roughly half-way between RS-1 and RS-2. 
Fig.~\ref{fig:RS1_RS2} shows the two remote stations in their final configuration,
with RS-2 to the left; note the displacement of the antennas from the electronics boxes, to suppress locally generated noise, relative to the 
original configuration of RS-1. The expected azimuthal angular resolution for reconstruction of distant point sources follows from the absolute timing resolution in each of the stations, which is achieved by counting the number of clock ticks of the local 200 MHz clock in each FPGA from the local pps (pulse per second) GPS signal (with a resolution of $\approx$10 ns) to the time of the trigger latch, which is therefore discretized in 5 ns bins. The overall timing resolution is therefore of order $\sqrt{2}\times\sqrt{(10~ns)^2+(5~ns)^2}$, or about 16 ns, corresponding to an angular resolution of order 3m/60m, or 3 degrees in azimuth. Although this is of the same order as the beam width, it is, nevertheless, sufficiently small to discriminate backgrounds arising from outside the effective signal volume.

A discussion of the front-end antenna, including design, modelling and calibration has been presented elsewhere\cite{Isaac14}, therefore
the remainder of this paper consists of a survey of the remaining components, in sequence, with performance benchmarks presented, where possible.

\subsection{Mixer Module} 
\label{section4.2}

\begin{figure}
\centerline{\mbox{\includegraphics[width=0.48\textwidth]{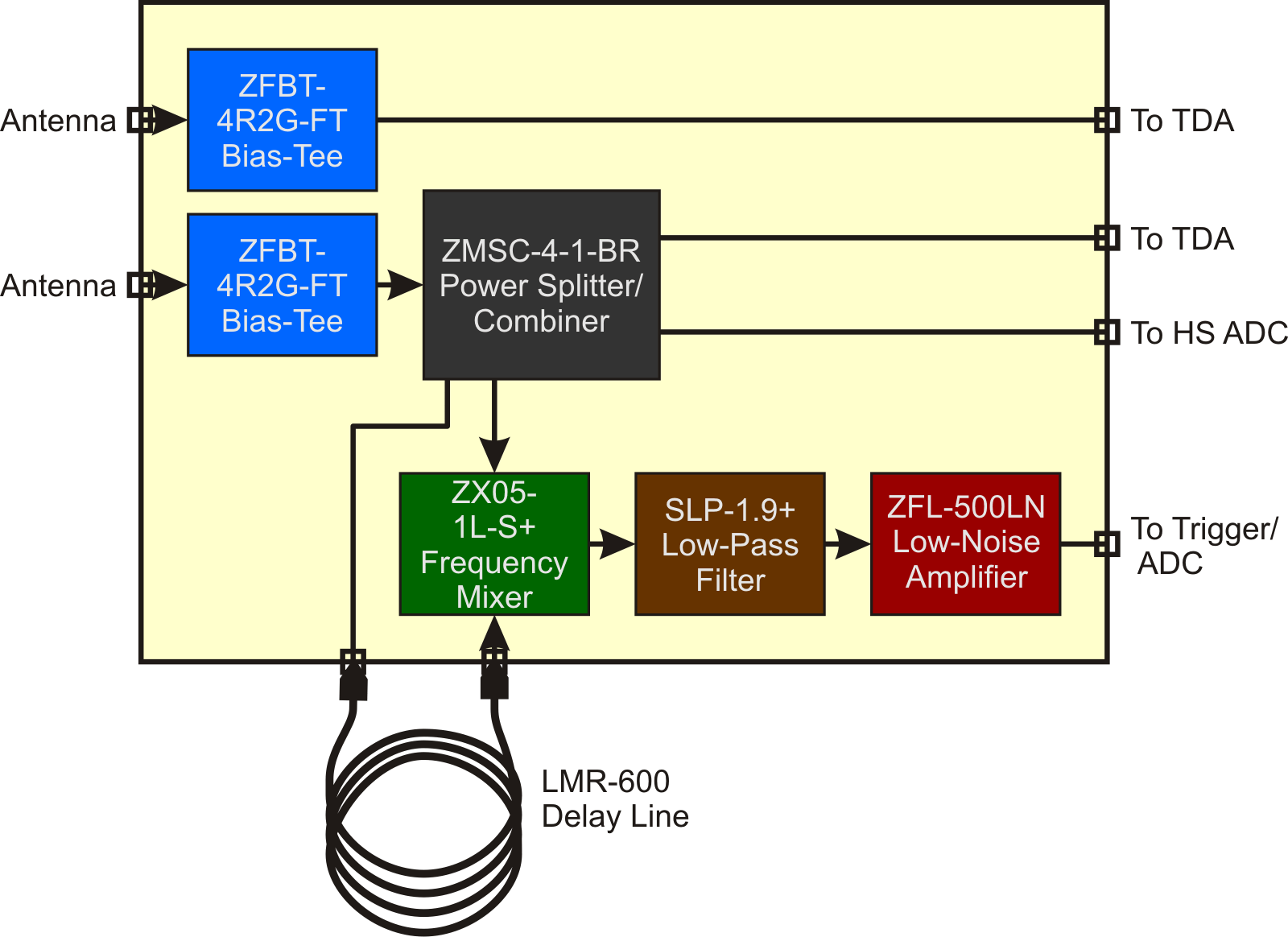}}}
\caption{%
      Block diagram showing the different components of the mixer module.
	\label{fig:mixer_module}}
\end{figure}

Before triggering, the raw input signals must be frequency-filtered and amplified in order to suppress known backgrounds (e.g., the known FM band) and also to ensure that the signal strength is comfortably within the dynamic range of the signal digitizer. The Mixer Module therefore includes splitters (ZMSC-4-1-BR, Mini-Circuits), filter (SLP-1.9+, Mini-Circuits), amplifier (ZFL-500LN+B, Mini-Circuits) and bias tees (ZFBT-4RG-FT, Mini-Circuits) as shown in Fig.~\ref{fig:mixer_module}. 

\begin{figure}
\begin{center}
\begin{minipage}{0.5\textwidth}
\centerline{\mbox{\includegraphics[width=0.98\textwidth]{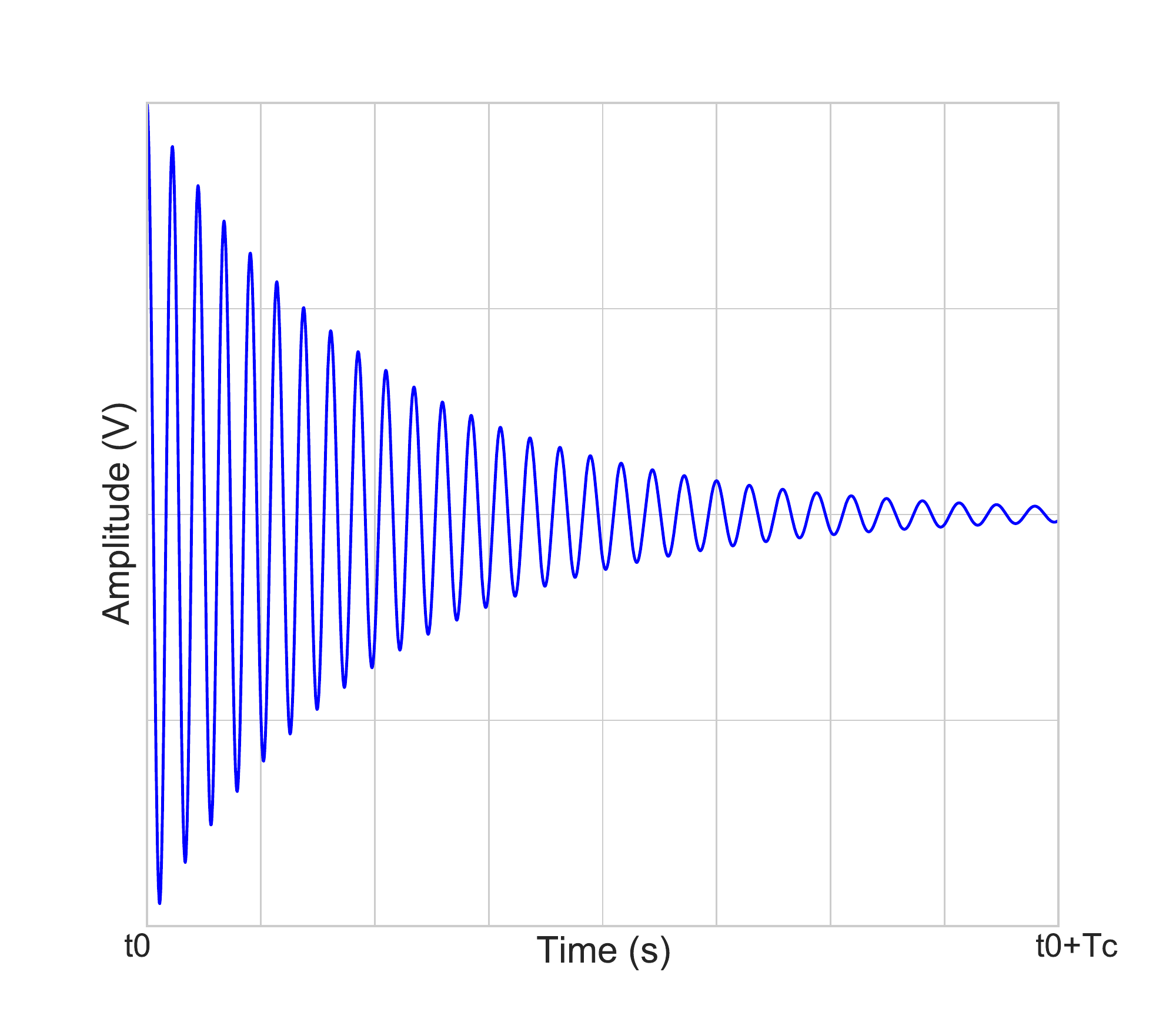}}}
\label{fig:chirp_demo_time}
\centerline{\mbox{\includegraphics[width=0.98\textwidth]{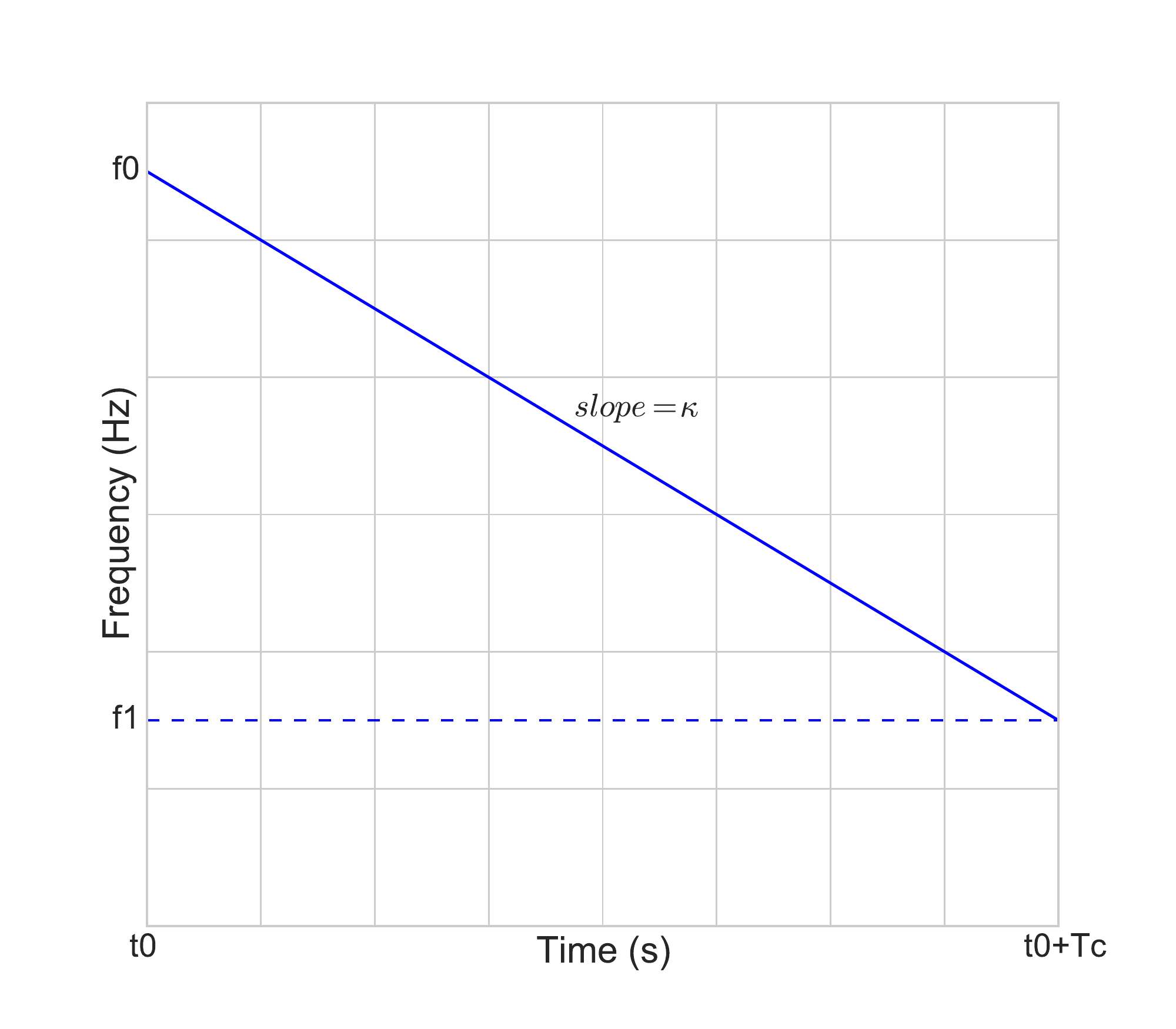}}}
\label{fig:chirp_demo_ft}
\caption{ A linear down-chirp with changing amplitude in the time domain (top) and frequency-time domain (bottom).
\label{fig:chirp_demo}}
\end{minipage}
\end{center}
\end{figure}

The mixer module also includes a Mini-Circuits ZX05-1L-S+ mixer, which is crucial to the identification of putative UHECR-induced radar echoes.
As detailed elsewhere\cite{Isaac14}, the EAS-reflected chirp signal can be characterized by a duration $T_c$ seconds, a time-varying Amplitude A(t) having with start phase $\phi_0$, start frequency $f_0$, stop frequency $f_1$ and chirp rate $\kappa$ Hz/s. Both up and down chirps are treated the same in the detector and have equal triggering efficiency. An example of a signal of interest is shown in Fig.~\ref{fig:chirp_demo}. Assuming that it is centered around time t=0, such a chirp signal has the form, in terms of the parameters above,

\begin{equation} \label{eq:chirp_equation}
s(t) = A(t)sin(\phi_0 + 2\pi f_0 t + \pi \kappa t^2)
\end{equation}

\subsubsection{De-chirping}
\begin{figure}
\includegraphics[width=0.48\textwidth]{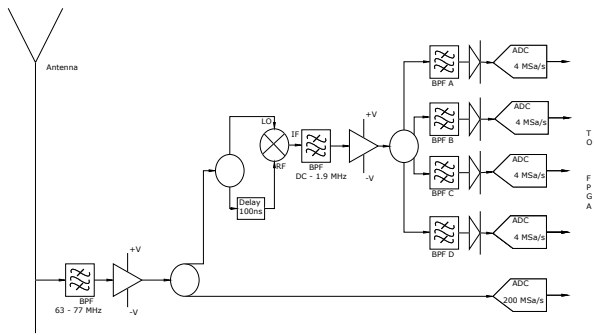}
\caption{Radar block diagram illustrating the down-conversion process along with filtering and enveloping.\label{fig:radar_blockdia}}
\end{figure}

We have employed an analog detection scheme (``de-chirping'') designed to 
detect the presence of the signal s(t) without prior knowledge of the chirp rate $\kappa$. In the first step, the signal is down-converted to a monotone.  To achieve this, the signal is mixed with a delayed copy of itself, i.e $s(t) \bigotimes s(t-\tau)$, with $\tau\sim$100 ns, as depicted in the radar block diagram in Fig.~\ref{fig:radar_blockdia}. 

\begin{figure}
\begin{minipage}{0.48\textwidth}
\flushleft{\mbox{\includegraphics[width=0.99\textwidth]{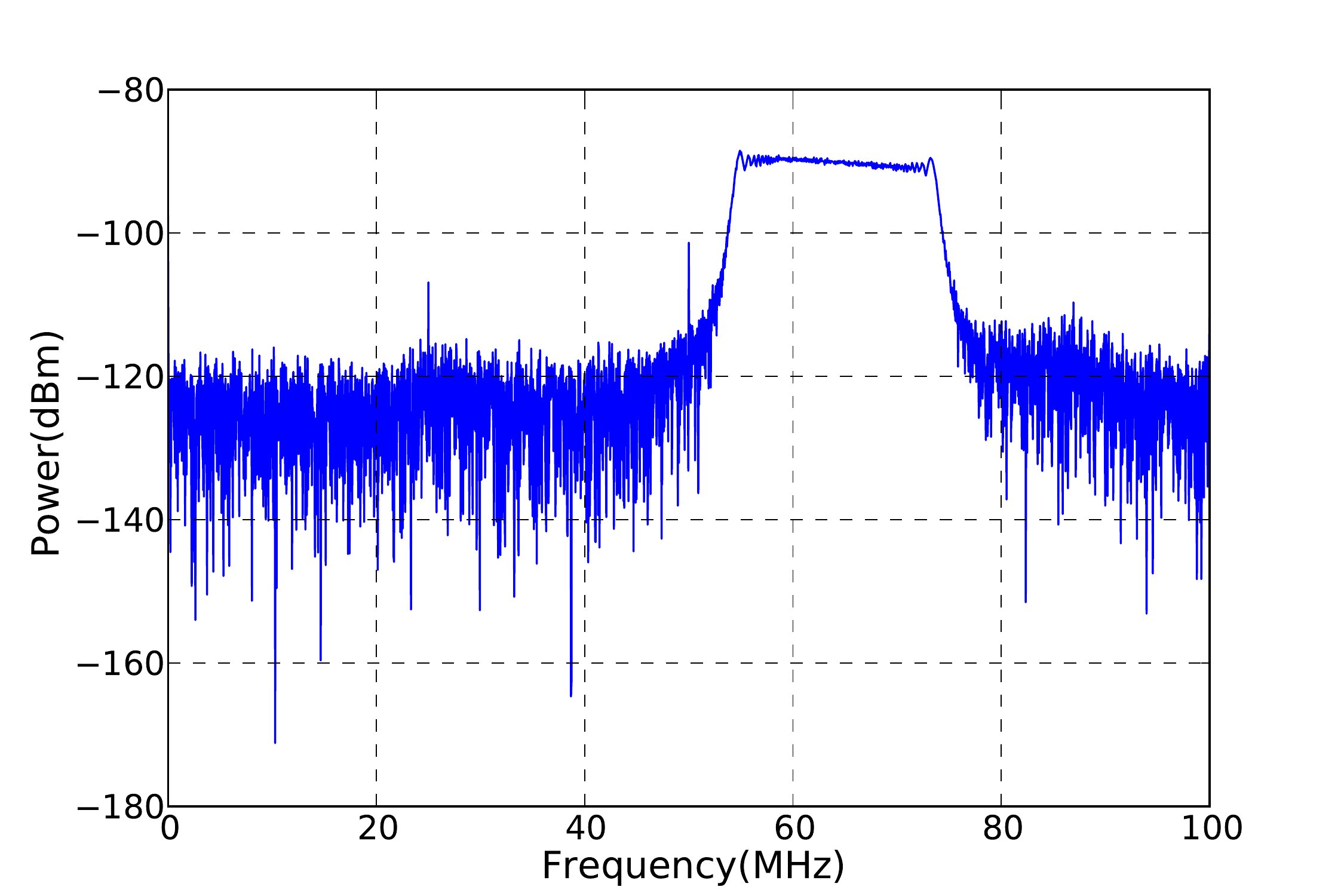}}}
\label{fig:chirp_fft}
\flushleft{\mbox{\includegraphics[width=0.99\textwidth]{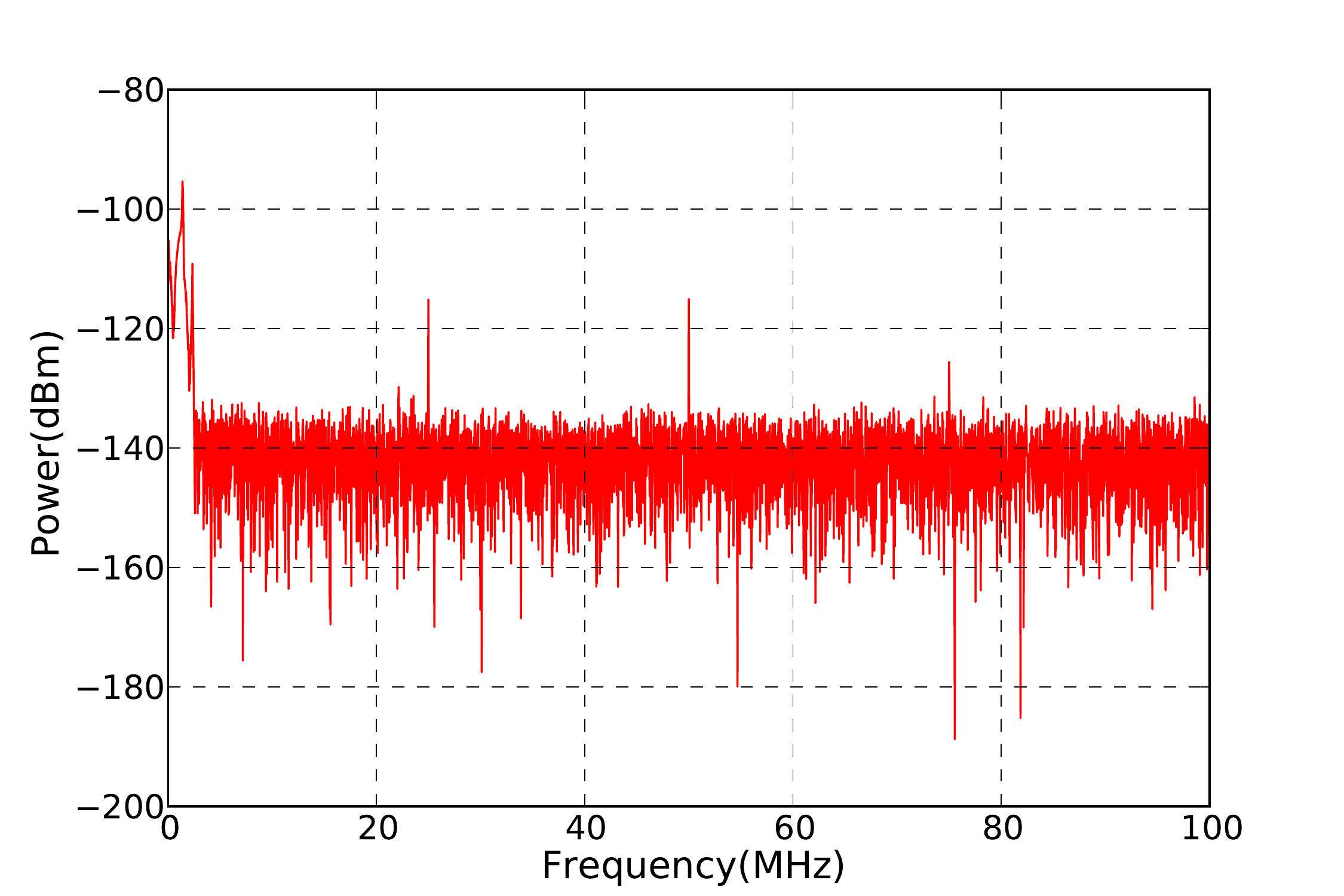}}}
\label{fig:dechirp_fft}
\caption{ Top: power spectrum of -10 MHz/$\mu s$ chirp created by a laboratory signal generator, prior to mixing. Bottom: power spectrum of monotone signal after signal mixing and low-pass filtering. The chirp is evident as the left-most peak in this distribution at $\sim$2 MHz. Higher-frequency overtones are suppressed in the triggering logic by bandpass filtering.
\label{fig:chirpanddechirp_fft}}
\end{minipage}
\end{figure}

For an incident chirp signal, the non-linear components in the mixer result in a product term that yields a monotone at a beat frequency 

\begin{equation} \label{eq:beat_equation}
f = \kappa\tau,
\end{equation}

where $\tau$ is the delay time. This delay is created with a short length ($\sim$30 m, selected as a length which results in destructive interference between the direct and time-delayed 54.1 MHz large-amplitude carrier signals, which would otherwise obscure the chirp) LMR-600 cable, which produces negligible losses.  This is illustrated in Fig.~\ref{fig:chirpanddechirp_fft}, in which a -10 MHz/$\mu s$ chirp is down-converted to a 1 MHz monotone. The 1 MHz monotone can then be detected using a simple threshold-crossing criterion, in which case a trigger is generated and waveform-capture initiated.

\subsection{Triggering Scheme} 
\label{section4.3}

The expected value of chirp slopes from EAS echoes are typically between  -10 to -1 MHz/$\mu$s. Consequently, and based on the previous discussion, a 100 ns time delay between the initial input signal and the delayed copy results in a down-converted signal with frequency between $\sim$100 kHz and 1 MHz. To trigger on such signals, it is advantageous to split the mixed signal into multiple copies, each of which is then passed through a distinct band-pass filter and a subsequent envelope detector. Different frequency bands are then compared by majority logic in an FPGA, requiring a trigger criterion of no more than two adjacent frequency bands above threshold. This trigger criterion is designed to suppress 
broadband impulsive noise, which will typically fire all four bands. 

\begin{figure}
\begin{center}
\begin{minipage}{0.5\textwidth}
\flushleft{\mbox{\includegraphics[trim=1.5cm 0.0cm 3.3cm 1.5cm,clip=true,width=0.98\textwidth]{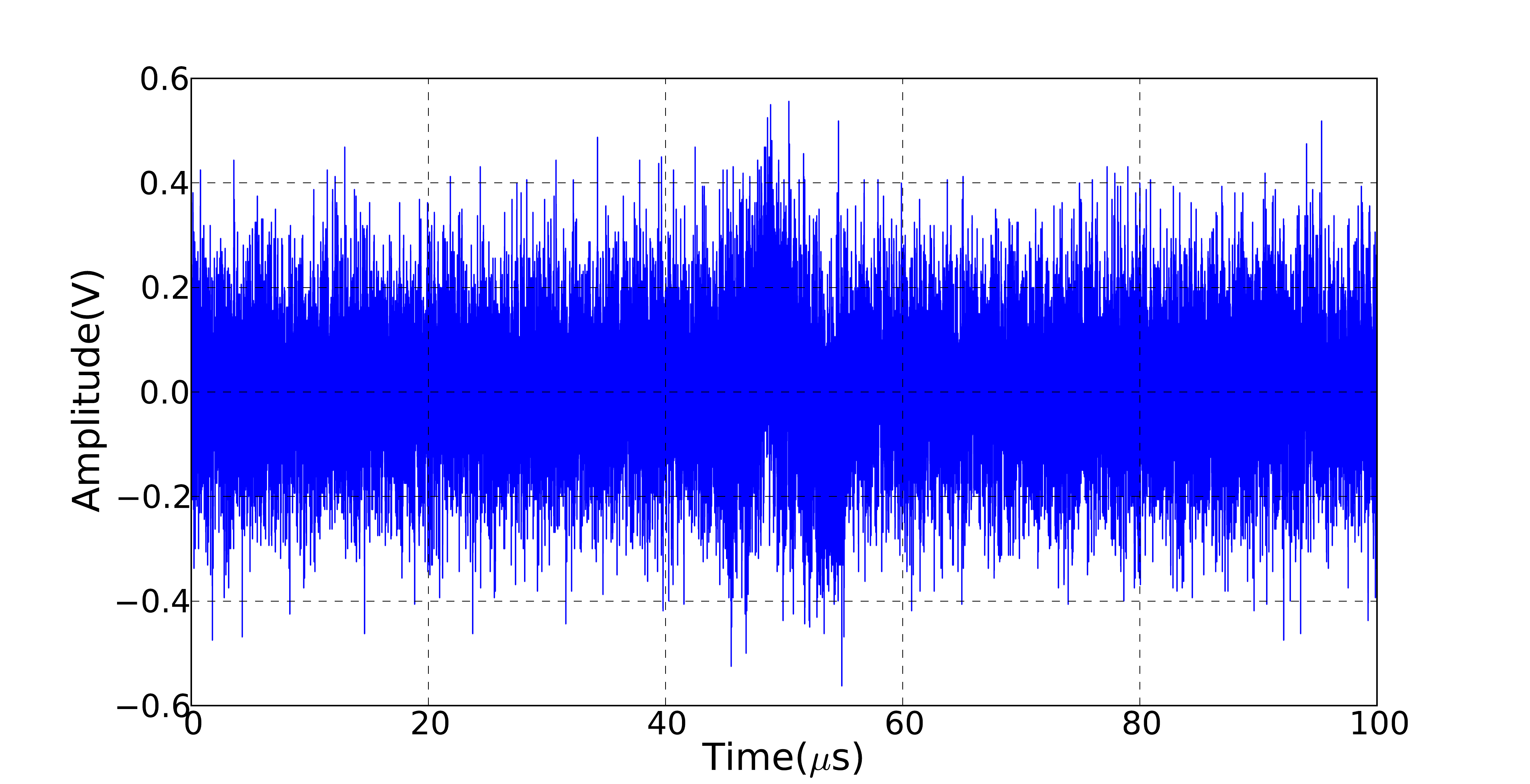}}}
\label{fig:rs_orig_td}
\flushleft{\mbox{\includegraphics[trim=1.3cm 0.0cm 3.3cm 1.5cm,clip=true,	width=0.98\textwidth]{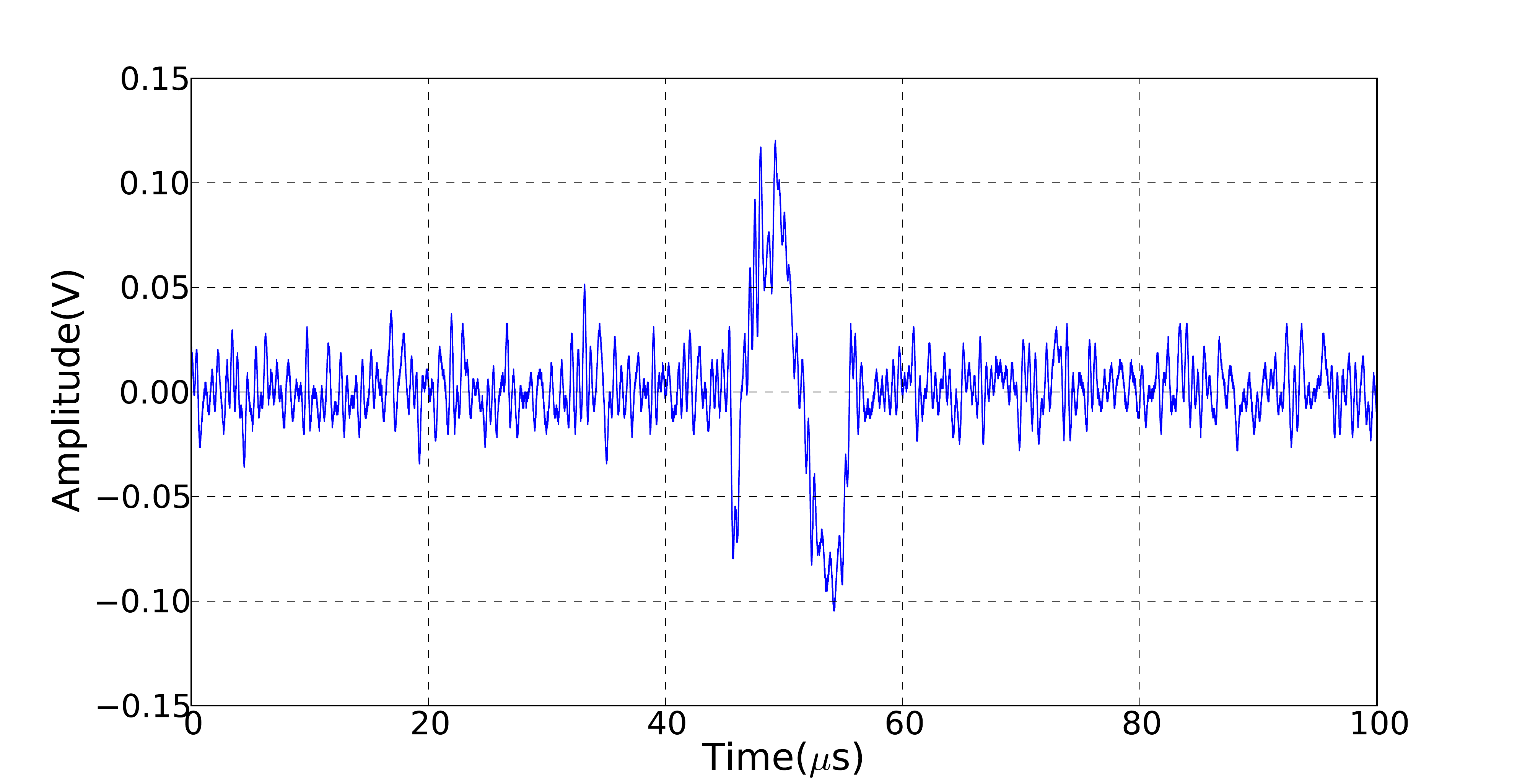}}}
\label{fig:rs_dc_td}
\flushleft{\mbox{\includegraphics[trim=0.8cm 0.0cm 3.3cm 1.5cm,clip=true,width=0.98\textwidth]{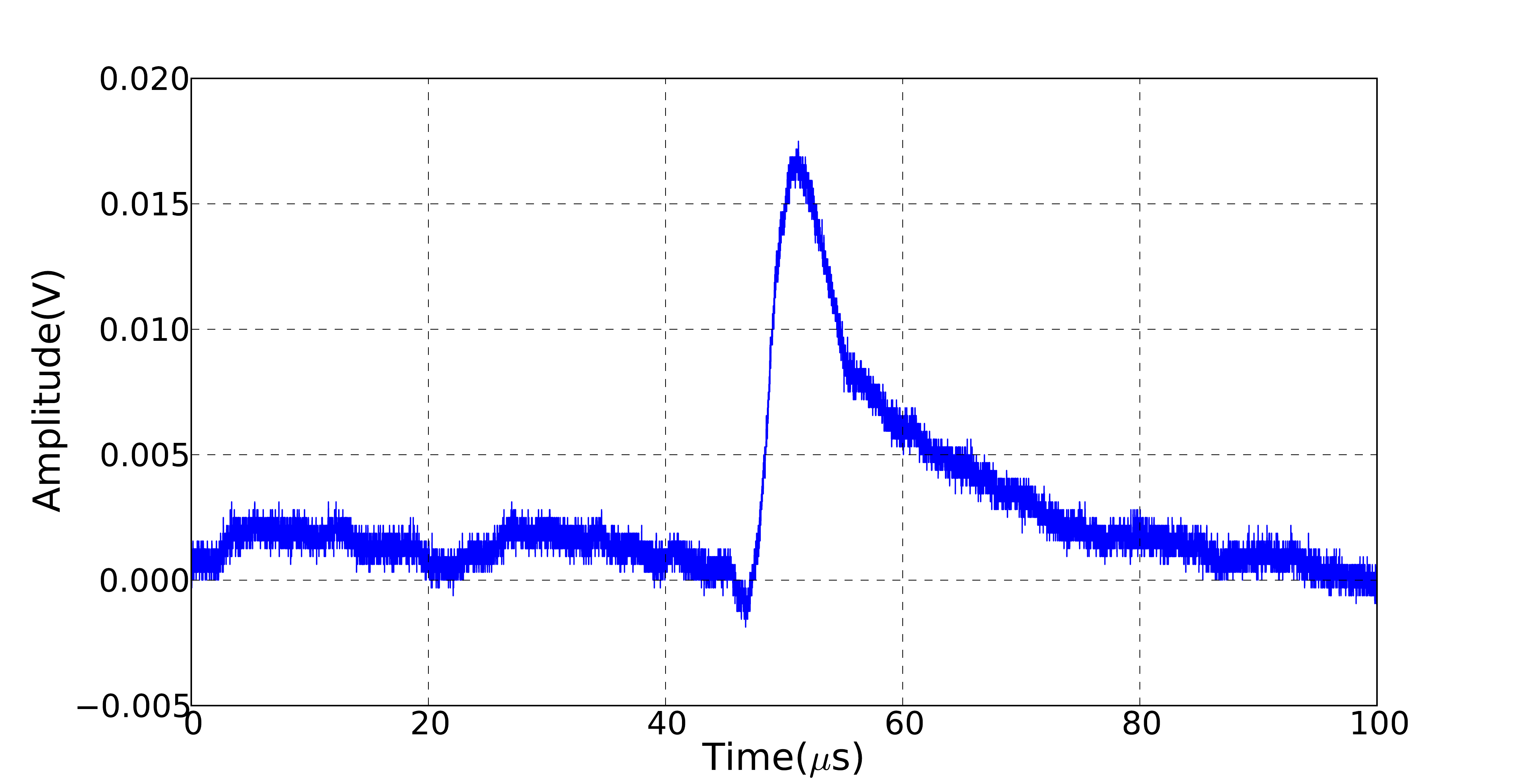}}}
\label{fig:rs_envdet_td}
\caption{Top: A 0 dB SNR and 1 MHz/$\mu$s chirp embedded in noise prior to de-chirping.  Middle: The monotone signal after input chirp is mixed with the delayed copy of itself and passed through a low-pass filter.  Bottom: Resulting monotone passed through the Agilent 8471D power detector.
\label{fig:rs_sig_chain}}
\end{minipage}
\end{center}
\end{figure}

 Fig.~\ref{fig:rs_sig_chain} (captured on an oscilloscope) illustrates the triggering scheme using a signal generated in the lab. After mixing and filtering, the signal is then passed through an envelope detector (8471D; Agilent, Inc.). Here, a chirp with 0 dB SNR with slope -1 MHz/$\mu$s is first band-pass filtered (41--100 MHz) and then amplified by 20 dB. The signal is then mixed and low-pass filtered (DC-1.9 MHz) and passed through the Agilent power detector.

\subsection{Chirp Acquisition Module (CAM)} 
\label{section4.4}

\begin{figure}
\centerline{\mbox{\includegraphics[width=0.48\textwidth]{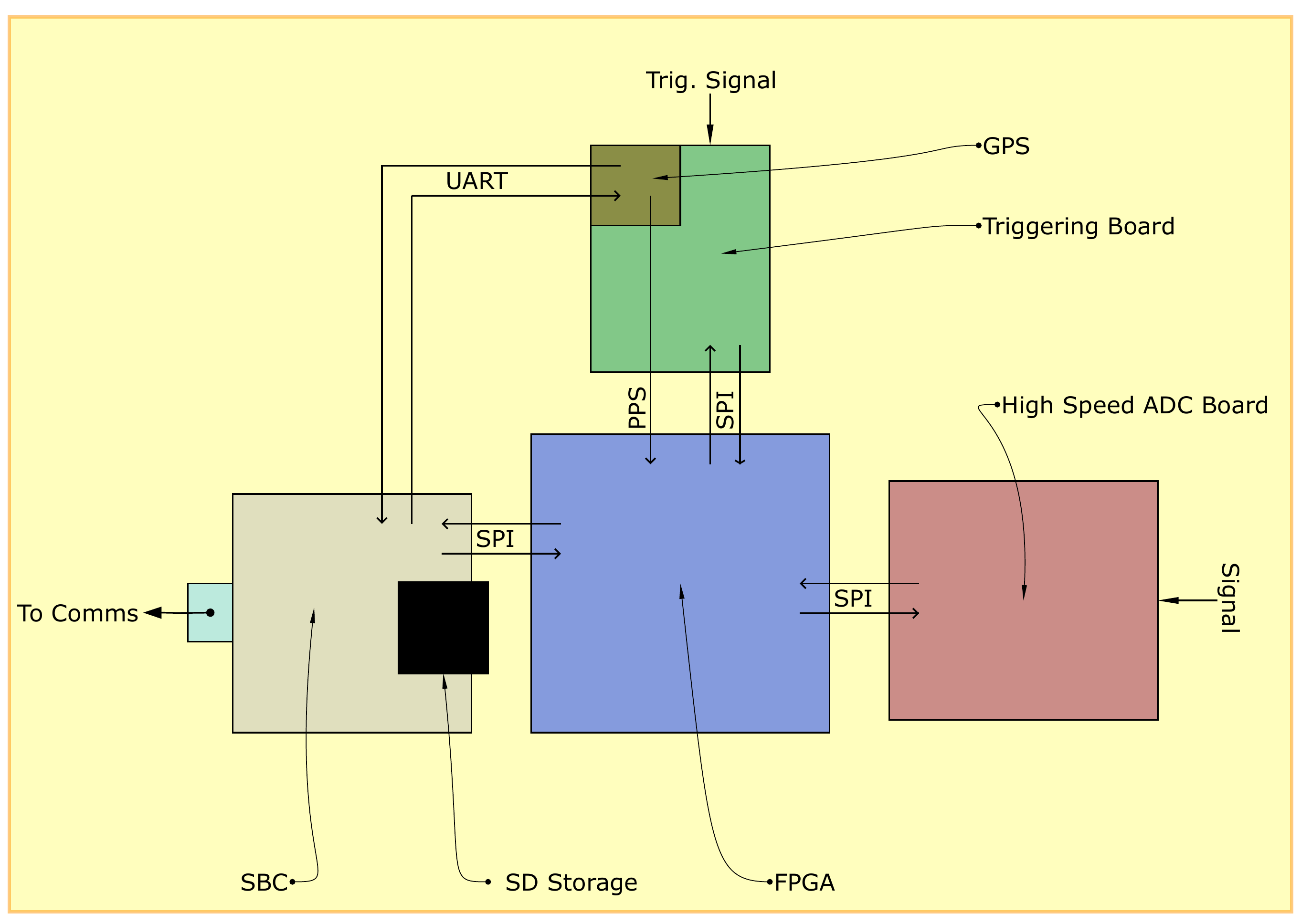}}}
\caption{Elements of the CAM unit illustrating the communications protocols. 
\label{fig:CAM_connections}}
\end{figure}

The CAM is an embedded system with a modular design that provides hardware and software integration for chirp detection and waveform capture. It is composed of five basic parts: a Trigger Board, High Speed Board, an FPGA, a GPS unit and a Single Board Computer (SBC), as shown in Fig.~\ref{fig:CAM_connections}. A description of each of these subsystems follows.

\subsubsection{Trigger Board}
\label{sub:Trigger_Board}
As the frequency of the de-chirped monotone is directly related to the signal chirp rate, that monotone is band-passed into four distinct frequency bands prior to discrimination. This is done to both ensure that the chirp rate is within a range commensurate with that expected for a cosmic ray, as well as to obtain a first estimate of the chirp rate itself.




A trigger signal is asserted only if the monotone power, as measured by the envelope detector, exceeds some adjustable, pre-set threshold $T_0$.
The envelope detector includes a rectifying diode, followed by an RC combination that allows the output waveform to follow the envelope of the signal.  
The output of the four band-pass filters are then passed into the ADC (ADC 80066, Analog Devices)\cite{AD80066}. The ADC is powered from a 5 V power supply, typically consuming 490 mW of power, and is packaged in a 28-lead Small-Shrink-Outline-Package (SSOP). The ADC is operated in the Sample and Hold Amplifier (SHA) mode with the 16-bit output multiplexed into 8-bit words and accessed in two read cycles clocking at 4 MHz per channel. 
The ADC register is programmed via the Serial Peripheral Interface (SPI) at 24 MHz.
Finally, the data are transferred via the 
FPGA pmod inputs for triggering (see Sec.~\ref{sub:fpga}).
Along the signal path, differential signaling is used to provide
superior common-mode noise rejection.

\subsubsection{High Speed Board}
\label{sub:High_Speed_Board}

The TARA High Speed Board is a commercially available Analog Devices AD9634 Evaluation Board\cite{AD9634}. The board, capable of sampling up to 250 MHz ADC with a total power consumption of 360 mW, affords 12-bit resolution. The ADC is clocked at 200 MHz and uses an 1.8V SPI port at 24 MHz for register programming and read back.

The 12 bit words are transferred using a custom adapter 
between the Evaluation Board's LVDS (Low Voltage Differential Signaling) parallel output port and the VHDC (Very High Density Cable) connector on the FPGA. 

\subsubsection{FPGA}
\label{sub:fpga}

We now provide details on the FPGA (Nexsys 3 digital system development board using the Xilinx Spartan 6 FPGA) timing, generated with an
FPGA board (Nexsys 3 using a Xilinx Spartan 6 FPGA, Digilent).
To configure the FPGA, the configuration file 
is first stored in a non-volatile parallel PCM (Phase Change Memory) device and then transferred to the FPGA on power-up using the BPI-UP port.  This is one of four possible configuration modes on the development board, and is achieved by removing all connections on the J8 jumper\cite{Nexsys3}.

The Nexys3 board includes a single 100 MHz CMOS oscillator connected to pin V10 (the
GCLK0 input in bank 2). PLL (Phase Locked Loop) and DCM (Digital Clock Manager) features on the board can be used to synthesize other frequencies. A Xilinx Clock Gen wizard 
is correspondingly used to synthesize a 48 MHz clock.
These are then further sub-divided to synthesize clocks for transfer and synchronization between peripheral boards, as well as other requisite functions.


The triggering board is connected to the FPGA development board via the pmod connector. These pmod connectors are 2x6 right-angle, 100-mil female connectors that mate with standard 2x6 pin headers, with each set of 12 pin pmod connectors comprising two 3.3V VCC signals, plus two ground signals and eight logic signals.

Words from the triggering board are de-serialized from the parallel input ports in the FPGA via the implementation of a Finite State Machine (FSM). Once shifted in, a comparator is invoked and the trigger logic queried.

In order to regularly monitor the environment, a ``forced'', or snapshot trigger is also implemented on 0xFF clock cycles of the GPS pps (pulse per second) clock. This corresponds to a typical time interval of 4.25 minutes between forced triggers.


All I/O’s to the  VHDC connector are routed as matched pairs to support LVDS signaling, powered at 2.5V; all FPGA pins routed to the connector are located in FPGA I/O bank0. There are 20 matched pairs of data signals, 20 ground signals and 8 power signals on the VHDC connector \cite{Nexsys3}. 

\begin{figure}
\centerline{\mbox{\includegraphics[width=0.48\textwidth]{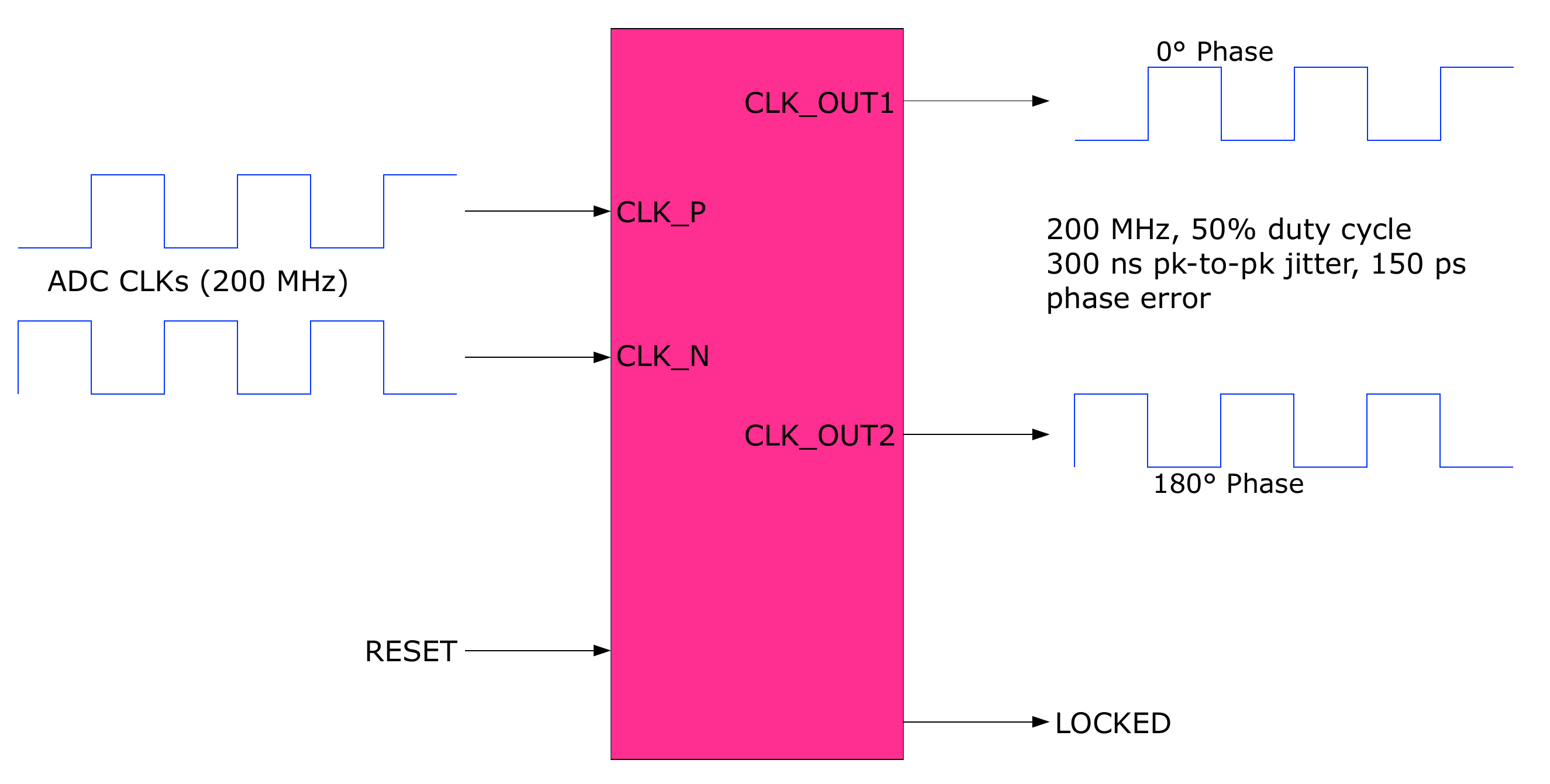}}}
\caption{High Speed Board's 200 MHz differential clocks and Xilinx ClockGen wizard used in synthesizing out-of-phase 200 MHz frequencies. 
\label{fig:differential_clocking}}
\end{figure}

The 200 MHz differential clock outputs from the High Speed Board are brought into the FPGA and used in frequency synthesis, via the Xilinx ClockGen Wizard, 
using two clocks at 200 MHz, aligned such that one is 180 degrees out of phase with the other (see Fig.~\ref{fig:differential_clocking}).

The AD9634 ADC sends even/odd bits on the rising/falling edge of the sampling clock. A DDR (Double Data Rate) interface \cite{XilinxUG381} is used between the FPGA and the ADC.
The bits are aligned per the C0 non-inverted clock, synthesized by use of an extra flip-flop.

IBUFDS (Input BUFfer Differential Signaling) is a differential I/O primitive \cite{XilinxUG381} that is instantiated for the signals from the ADC and have two pins, corresponding to the P and N channels of the differential signal.

\begin{figure}
\centerline{\mbox{\includegraphics[width=0.48\textwidth]{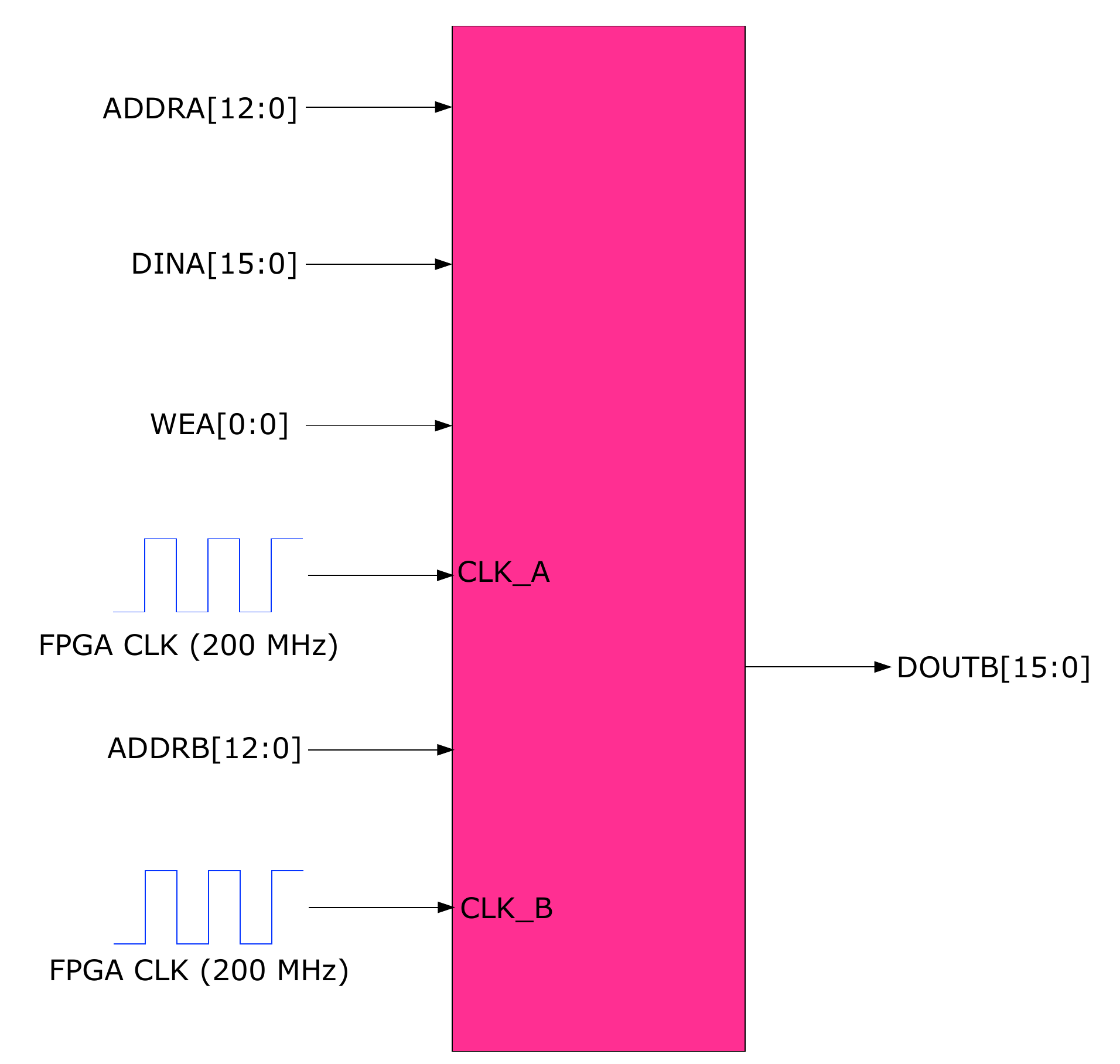}}}
\caption{Xilinx Block Memory Generator Signals.
\label{fig:circular_buffer}}
\end{figure}

Bits are then serialized via an FSM and stored in a circular buffer with a RAM depth of $2^{13}$ (8192) words and width of 16 bits (the actual width is 12 bits but the buffer is padded with four bits; bits 0,7,8,15 are redundant) that are continuously written and read at 200 MSa/s. The Xilinx Block Memory Generator Wizard 
handles the writing, reading and memory management, employing one 9K BRAM and seven 18K BRAMs (see Fig.~\ref{fig:circular_buffer}).

For each trigger, the 16 bit words ($2^{13}$ [8192] bit depth) are written into a FIFO at 200 MSa/s and then read out at 8 MSa/s (SBC's SPI clock). 

In addition to the signal captured by the FPGA from the High Speed Board, the triggering information is also saved. This information forms the header of the data to be transferred. 

\begin{figure}[!hb]
\centerline{\mbox{\includegraphics[width=0.48\textwidth]{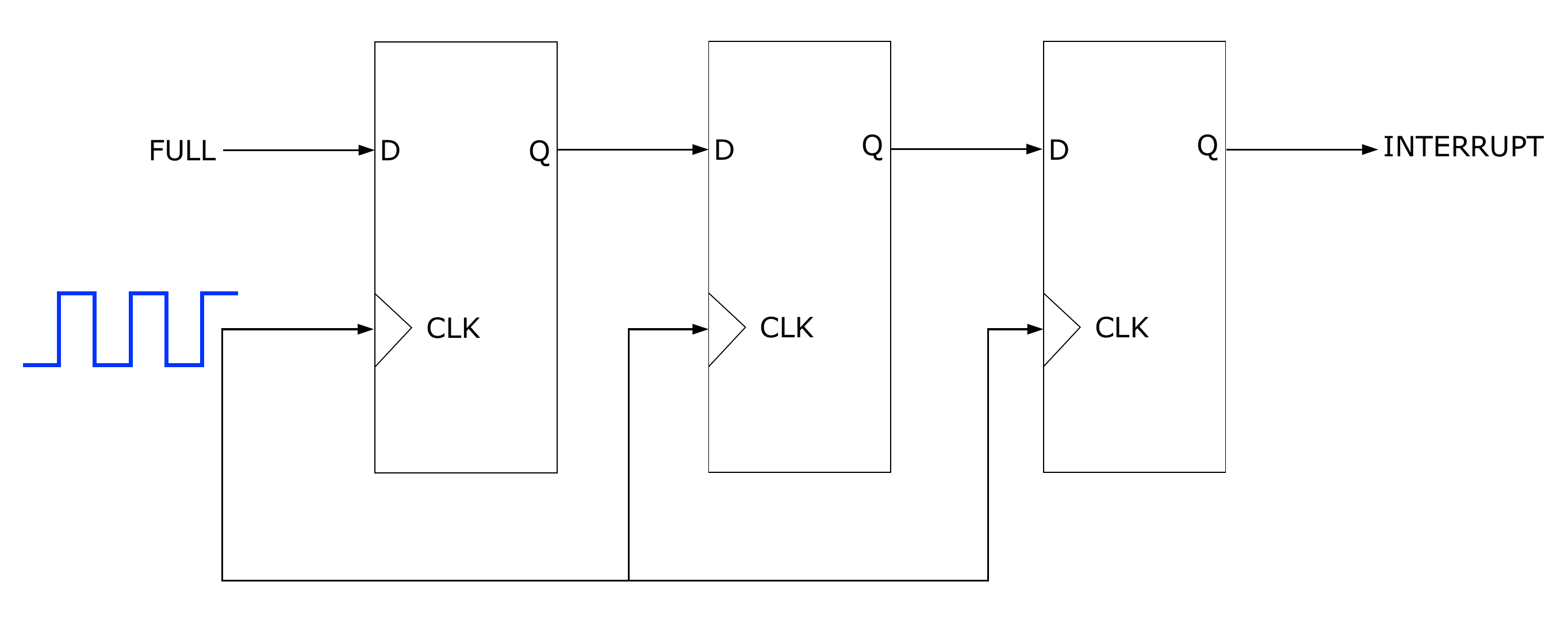}}}
\caption{Implementation of D Flip Flop to assert an Interrupt signal.
\label{fig:dflipflop}}
\end{figure}

The threshold level for triggering is set via a Raspberry Pi\cite{RaspPi} SBC (described below) using a chip select line; the aforementioned data to the SBC are transferred using the same data line, but a different chip select is activated once an interrupt is initiated on the FPGA side. This interrupt is ideally initiated when the FIFO buffer is full; however, due to latencies arising due to the crossing of clock domains in the FIFO, three D synchronous flip-flops, as shown in Fig.~\ref{fig:dflipflop}, are instantiated after the assertion of ``full'' on the FIFO, and therefore before the interrupt is asserted.  An FSM controls the transfer logic between the FPGA and SBC.

\begin{figure}
\centerline{\mbox{\includegraphics[width=0.48\textwidth]{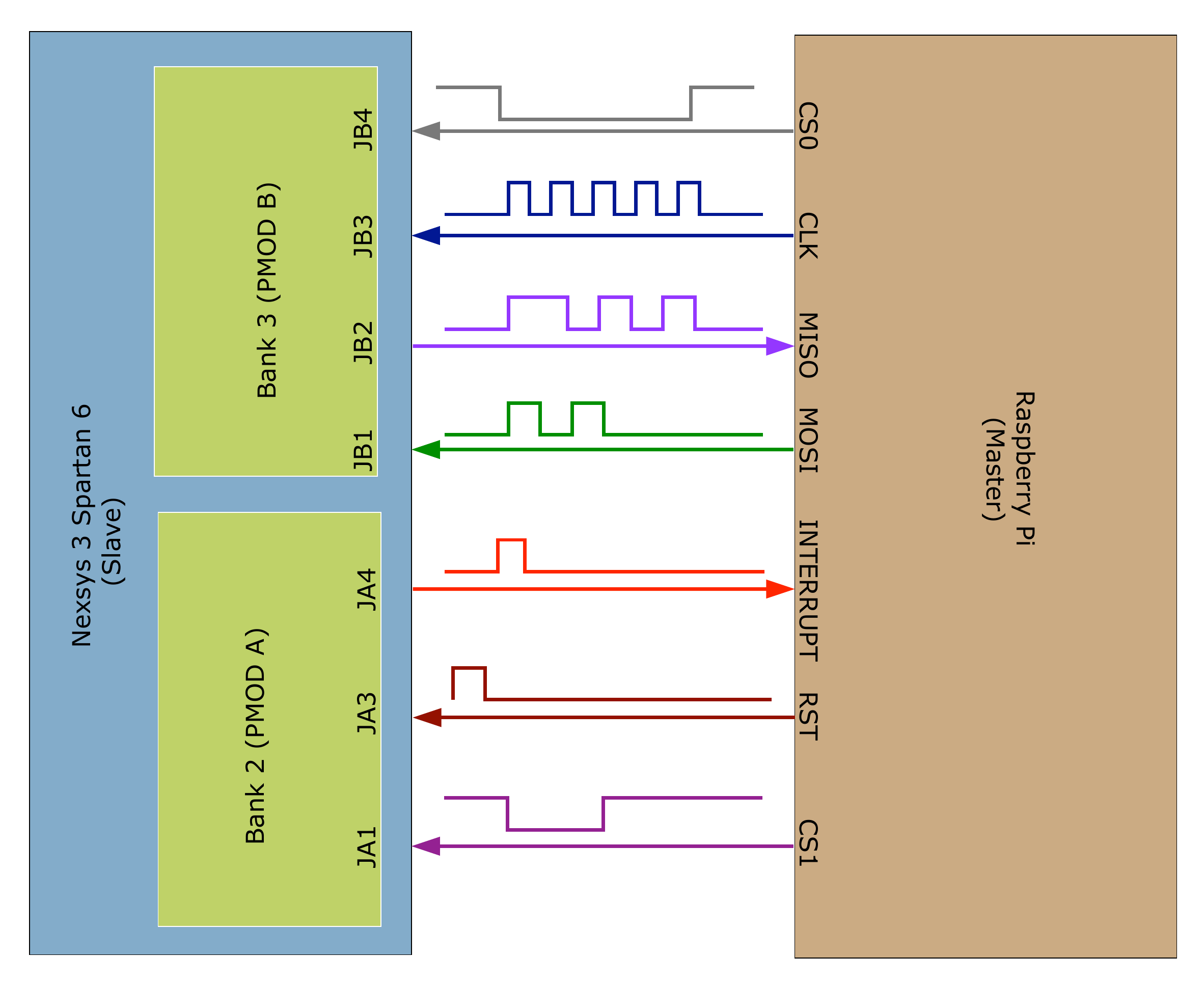}}}
\caption{The FPGA-SBC Interface.
\label{fig:fpgatfsbc}}
\end{figure}

In all transfers to/from the SBC(Master) and FPGA(Slave), serial peripheral interfaces are used and clocked at 8 MHz (see Fig.~\ref{fig:fpgatfsbc}).

\subsubsection{GPS}
\label{sub:GPS}

\begin{figure}[!h]
\centerline{\mbox{\includegraphics[width=0.48\textwidth]{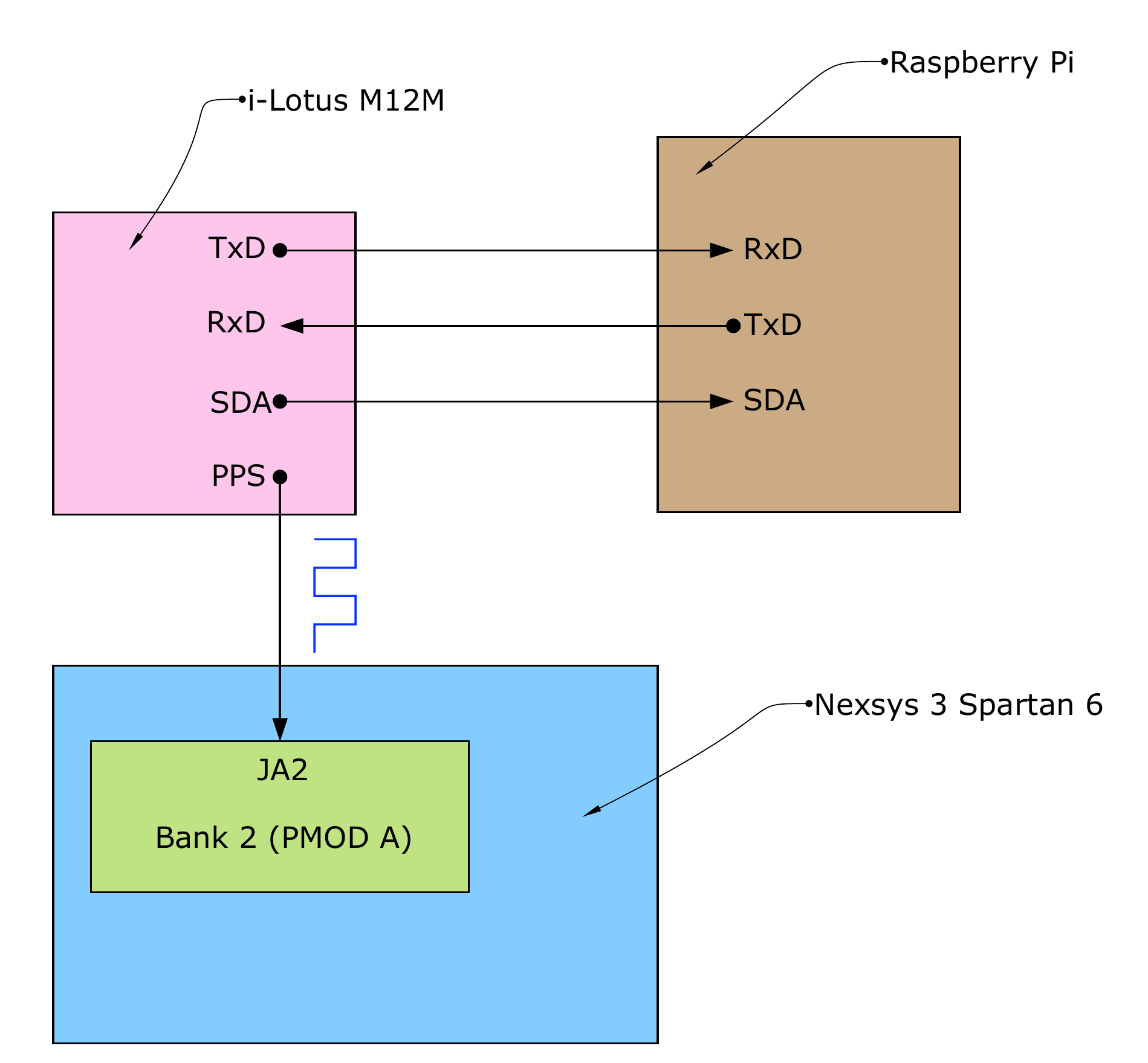}}}
\caption{GPS to FPGA and SBC Interface.
\label{fig:GPS_comms}}
\end{figure}

The GPS unit (i-Lotus M12M)\cite{ilotus} for event time-stamping is mounted on the Triggering Board for stability and convenient powering. 
The UART communications protocol is used to transfer data from/to the SBC. The GPS PPS (Pulse Per Second) is used as a counter to obtain forced triggers as described earlier in Sec. \ref{sub:fpga}. The functional Block diagram of the GPS to SBC and FPGA connections are shown in Fig.~\ref{fig:GPS_comms}.
Timing information is periodically queried from the GPS unit and used to update the time base on the SBC.

\subsubsection{Single Board Computer (SBC)}
\label{sub:SBC}
The Single Board Computer used in the CAM is the commercially available Raspberry Pi Model B Rev. 2 (RPi - B rev 2.).  The RPi uses the Broadcom BCM2835 SoC (System on Chip) that includes the ARM1176JZF-S 700 MHz processor, VideoCore IV GPU, 512 MB RAM and access to other I/O peripherals, as well as a separate three port USB Hub. 

\begin{figure}
\centerline{\mbox{\includegraphics[width=0.48\textwidth]{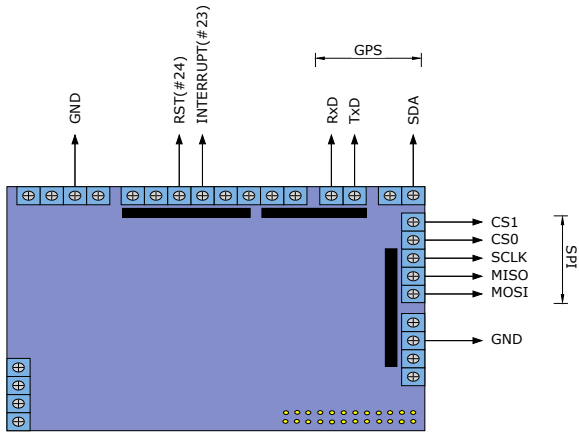}}}
\caption{Ardafruit Prototyping Pi Plate showing the corresponding GPIO pins to the FPGA and GPS units.
\label{fig:sbc_pins}}
\end{figure}

The RPi has a 26 ($2\times13$) pin 2.54 mm expansion header with 8 GPIO (General Purpose Input/Output) pins and dedicated peripherals such as SPI, and UART, along with 3.3V, 5V and GND supply lines. 
The board includes a 32 GB SD card to which data are stored and also where a minimal Raspbian Wheezy (118 MB image) is compiled with hard float support (3.6.11+ hardfp kernel). WiringPi\cite{WiringPi} written in C for the BCM2835 provides access to the GPIO and the Adafruit Prototyping Pi Plate provides convenient pin access to connect with the FPGA and GPS unit (see Fig.~\ref{fig:sbc_pins}).

\subsection{System Monitoring and Powering} 
\label{section4.5}

In addition to the CAM, the Chirp Detection Station includes several components to power the station and monitor environmental variables. These components are the Transient Detector Apparatus (TDA) Board, the current voltage and temperature (IVT) board and the system health monitor (SHM). A prototype station comprising just these components had been deployed in the Summer of 2013 to acquire environmental and powering information. In the revised stations, deployed in 2014, modifications were made and the system was updated to operate in conjunction with the chirp acquisition module. 

\subsubsection{TDA (Transient Detector Apparatus) Board}
\begin{figure}[!h]
\centerline{\mbox{\includegraphics[width=0.48\textwidth]{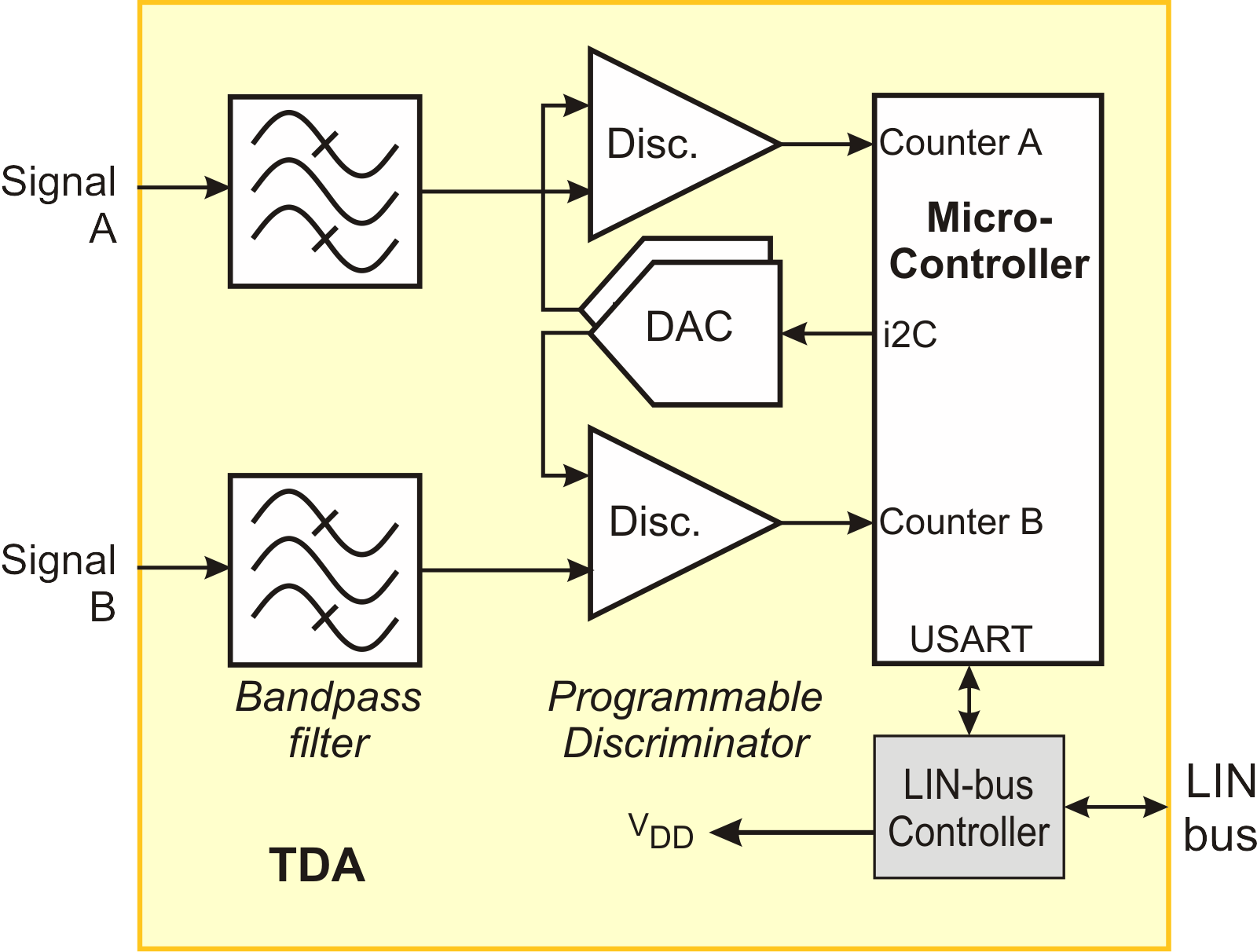}}}
\caption{Transient Detector Apparatus (TDA).
\label{fig:TDA_block}}
\end{figure}

The TDA continuously monitors ambient noise at the site by counting threshold crossings. The thresholds are set with Digital-to-Analog Converters that are remotely controlled through the System Health Monitor (SHM). The measurement period, usually 10 seconds, is also controlled from the SHM.
Prior versions, including that deployed in the prototype station, had a front-end amplifier in addition to a log amplifier, however both have been removed in subsequent iterations, and amplification now resides in the Antenna front-end. A functional block diagram of the TDA is shown in Fig.~\ref{fig:TDA_block}. A serial network protocol (LIN [``Local Interconnect Network''] - bus)
connects the TDA to the SHM.

\subsubsection{IVT (Current, Voltage and Temperature) Measurement Board}

\begin{figure}[!h]
\centerline{\mbox{\includegraphics[width=0.48\textwidth]{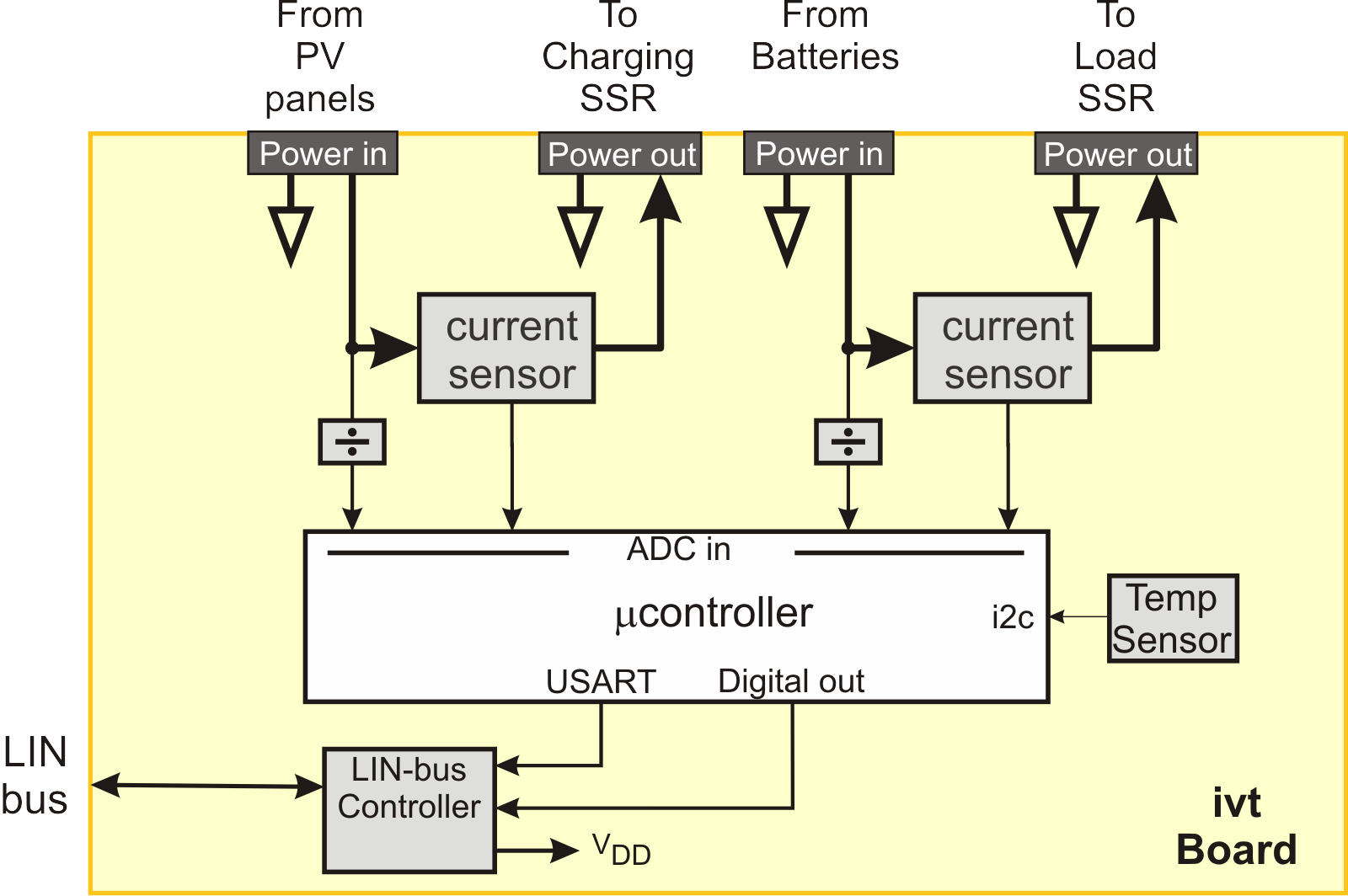}}}
\caption{Current Voltage Temperature (IVT) Board. 
\label{fig:ivt_block}}
\end{figure}

The IVT board consists primarily of current sensors, allowing voltage measurement and monitoring of the PV (Photo Voltaic) panels. It has the capability of measuring current and voltage for a wind turbine, however this has not been installed at present. In addition, the board includes temperature sensors along with the current and voltage sensors, all of which feed into a micro-controller. A functional block diagram of the IVT board is shown in Fig.~\ref{fig:ivt_block}. As in the TDA, a LIN-bus connects the IVT board to the SHM.

\subsubsection{SHM}



\begin{figure}[!h]
\centerline{\mbox{\includegraphics[width=0.48\textwidth]{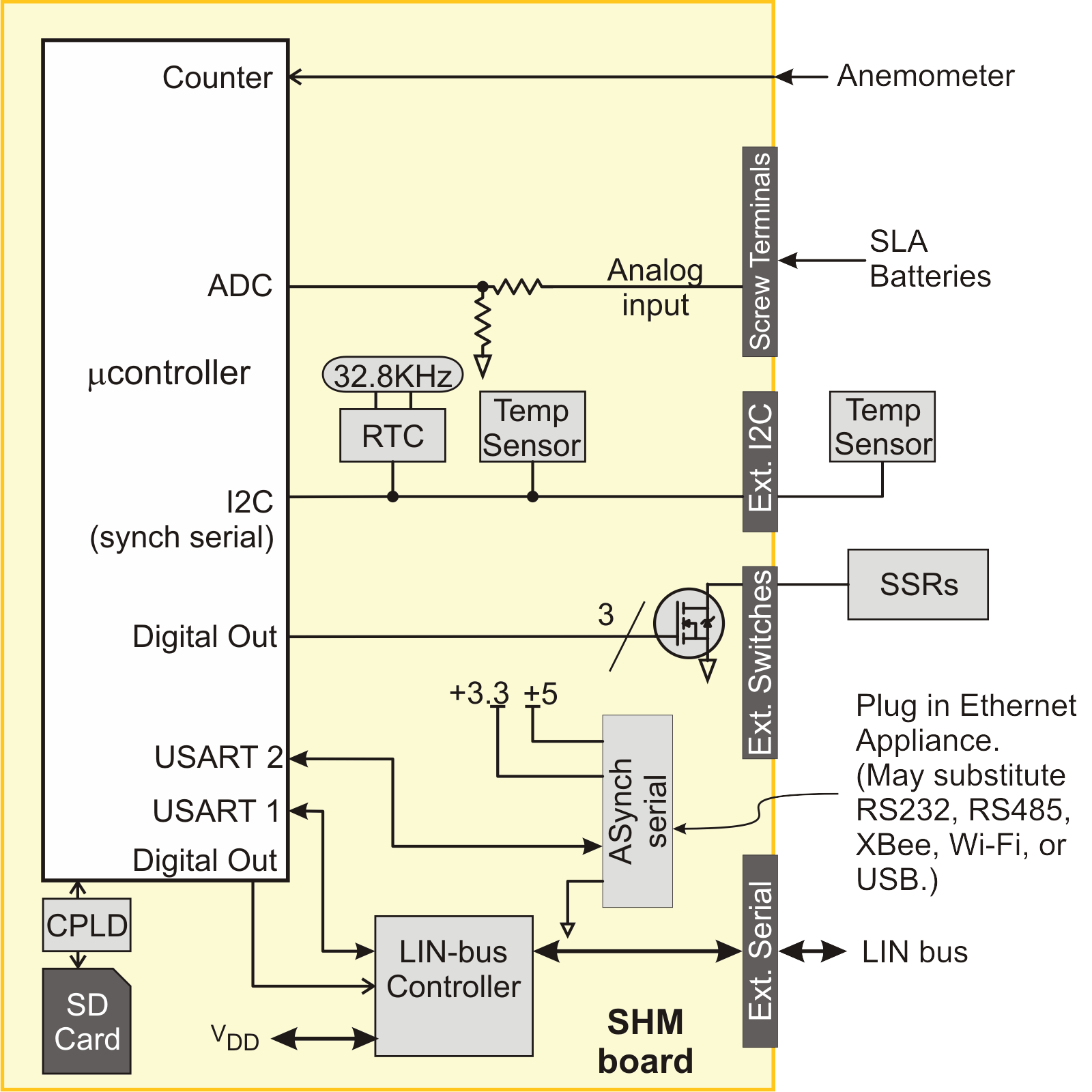}}}
\caption{System Health Monitor (SHM). 
\label{fig:SHM_block}}
\end{figure}

The System Health Monitor reports on the environmental and operational status of the Chirp Detector. This consists of a LIN bus connection to the TDA and IVT board, where ambient noise, PV current and voltage are measured. Additional temperature sensors measure the temperature both inside and outside the detector. The SSR's (Solid State Relays) are controlled via the SHM; using an Ethernet appliance, the status of the detector is reported on a web page. Operational characteristics of the detector are controlled via a CPLD (Complex Programmable Logic Device) with the environmental data stored onto an SD card.  A functional block diagram of the SHM is shown in Fig.~\ref{fig:SHM_block}.

\begin{figure}[!h]
\centerline{\mbox{\includegraphics[width=0.48\textwidth]{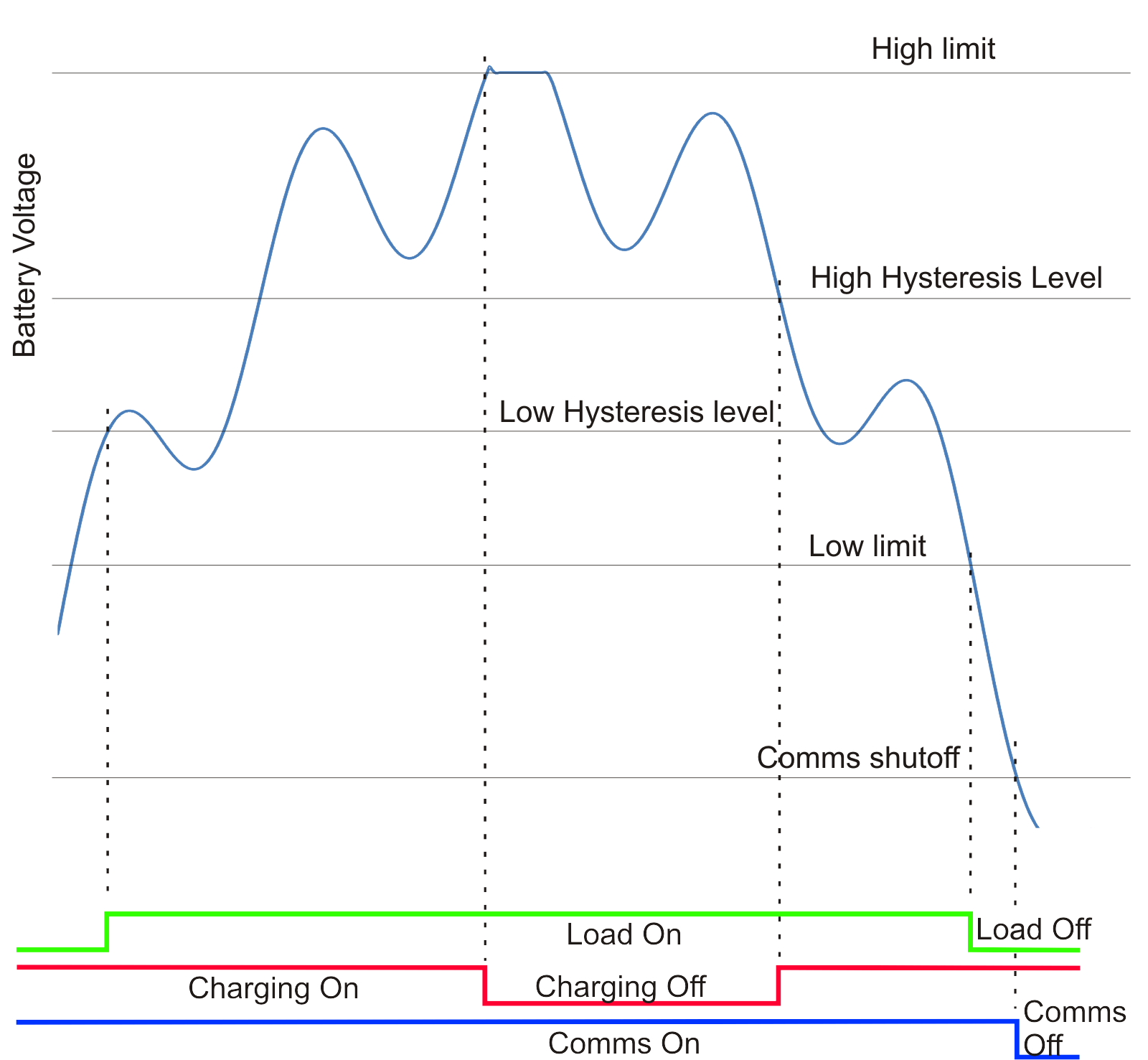}}}
\caption{Threshold Settings on the System Health Monitor (SHM).
\label{fig:SHM_Th}}
\end{figure}

The SHM has five adjustable threshold levels that dictate when communications, 
loads and charging are turned on/off (Fig.~\ref{fig:SHM_Th}).
To avoid over-charging, e.g., there is a failsafe such that a low-resistance, and therefore
high-dissipation load is swapped into the system in the unlikely event that the
system power demand is unable to draw normal charge from the batteries.

\subsection{Chirp Calibration Unit (CCU)}
\begin{figure}[!h]
\centerline{\mbox{\includegraphics[width=0.48\textwidth]{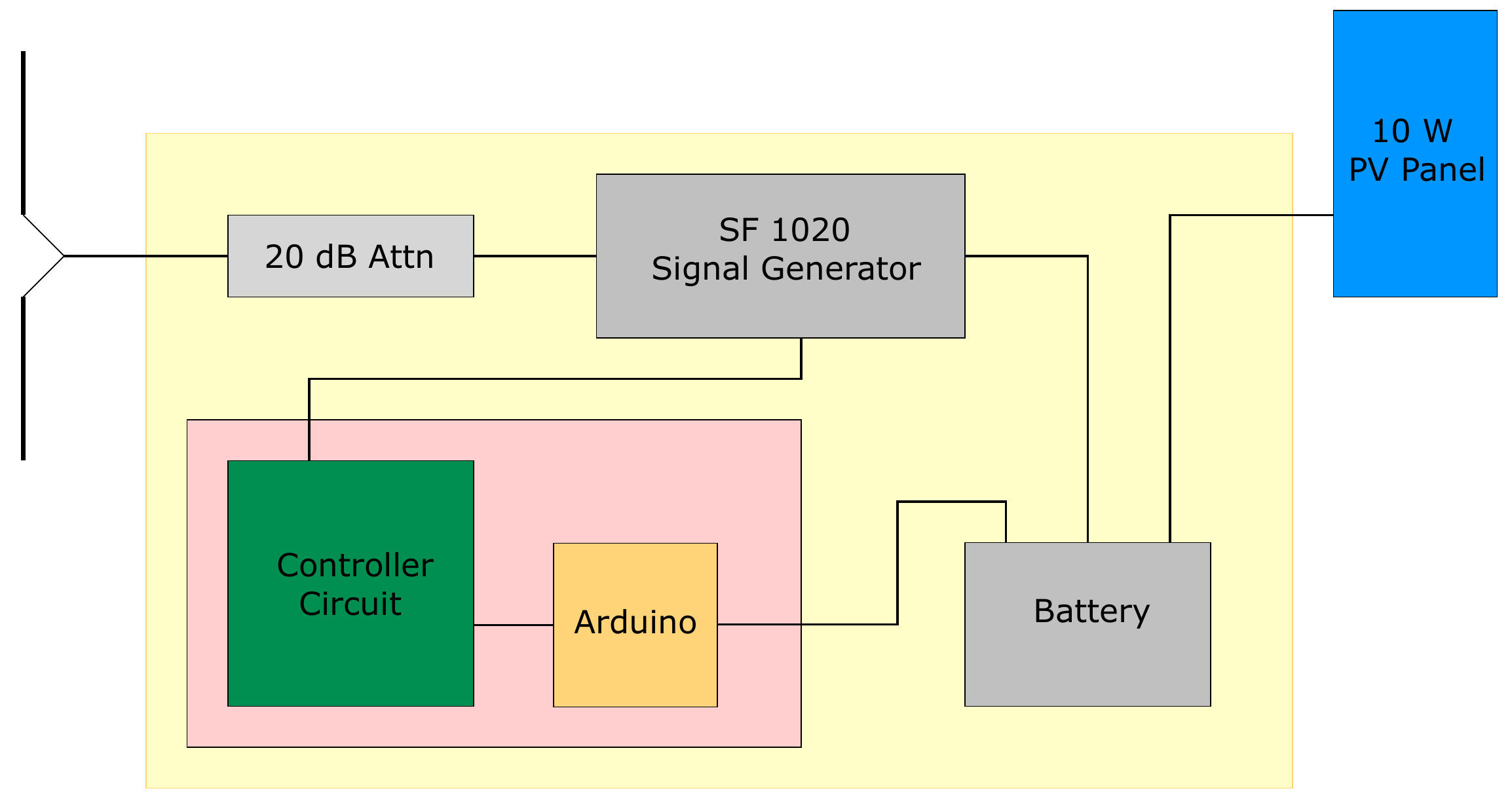}}}
\caption{Schematic of the Chirp Calibration Unit (CCU).
\label{fig:ccu}}
\end{figure}

In order to calibrate the remote stations independently, a Chirp Calibration Unit, with signal characteristics similar to those of an expected cosmic-ray induced chirp, has been deployed in the field. This unit is depicted in Fig.~\ref{fig:ccu}, and comprises a 'fat - dipole' transmitter antenna plus an Arduino controlling an SSR for the Signal Forge sf1020 signal generator.

\begin{figure}[h]
\centerline{\mbox{\includegraphics[width=0.48\textwidth]{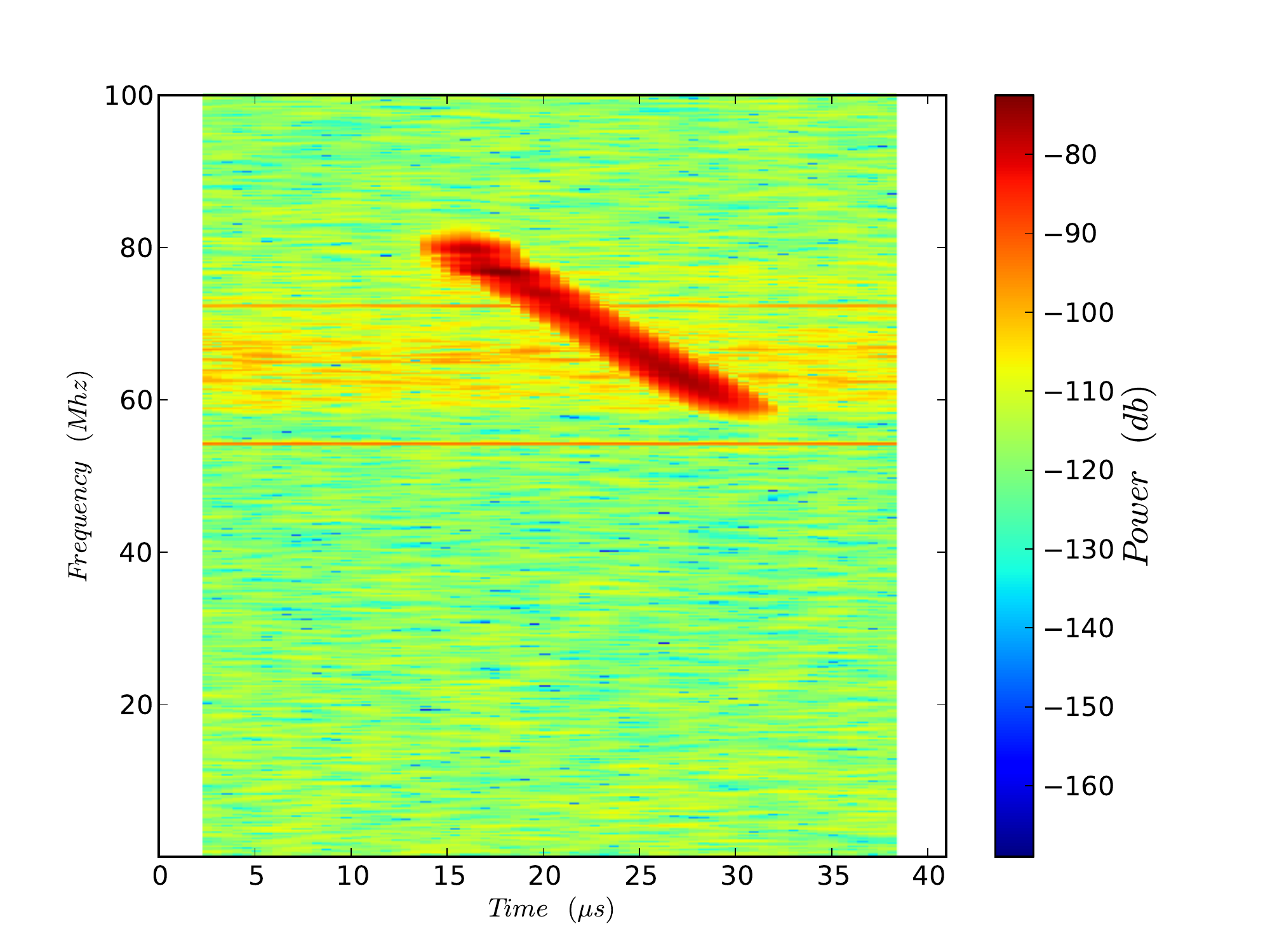}}}
\caption{CCU-triggered chirp as detected by RS-2. 
\label{fig:chirp_from_CCU}}
\end{figure}

The dipole was initially tuned to a frequency of 70 MHz, however, in order to transmit a broadband chirp-like signal, the radius of the dipole was re-optimized to efficiently transmit at 55 MHz. Stable 1 pps chirps spanning 80 MHz$\to$50 MHz and $20\mu$s duration over a period of 10 seconds are generated every 2 hours. 
The calibration unit was placed $\sim$45 m from the two remote stations to give high SNR chirp signals (20 dB SNR). A spectrogram of a calibration chirp is shown in
Fig.~\ref{fig:chirp_from_CCU}

\section{Deployment and Initial Performance}
\begin{figure}[!h]
\centerline{\mbox{\includegraphics[width=0.48\textwidth]{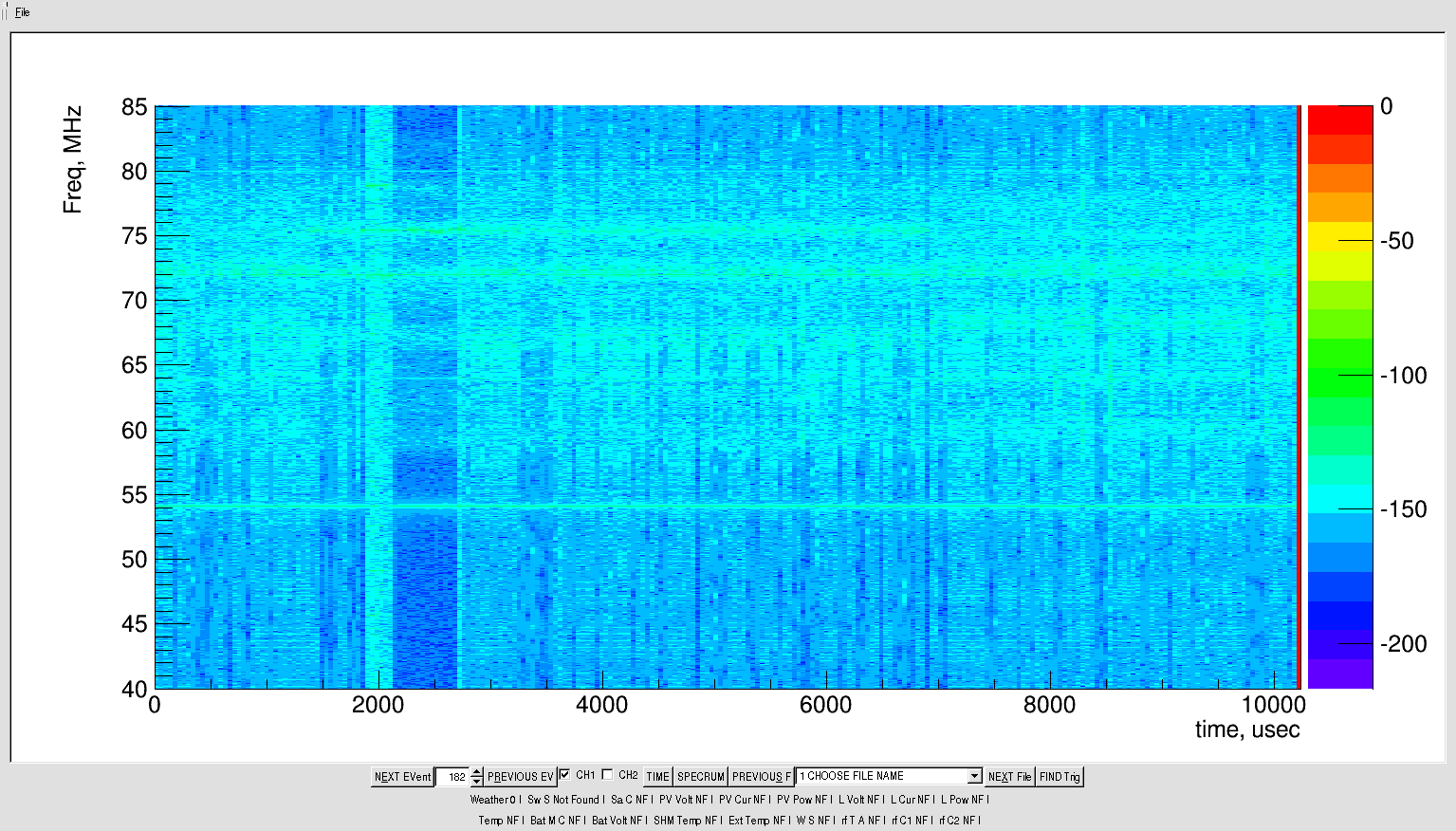}}}
\caption{Spectrogram of initial RS-1 data; data taken one day after initial installation (June, 2014).
\label{fig:EvViewer.png}}
\end{figure}

The initial deployment of RS-1 in June, 2014 was followed by an upgrade of RS-1 and also field deployment of RS-2 in November, 2014, followed by field installation of the CCU in January, 2015. The spectrogram for the first data taken by RS-1 is shown in Figure \ref{fig:EvViewer.png}. The two strongest and most persistent lines in the plot are those at 54.1 MHz (due to the carrier signal from the TARA transmitter) and one at 72 MHz which was later traced to a source in the RS-1 electronics and easily mitigated by physically moving the antenna away from the RS-1 electronics power box. The former was additionally suppressed by installation of tuned, in-line analog RLC filters in August, 2014. Absent any filtering, the 54.1 MHz carrier can saturate our amplifiers, resulting in an overall increase of the observed noise floor over tens of MHz.

\begin{figure}[!h]
\centerline{\mbox{\includegraphics[width=0.48\textwidth]{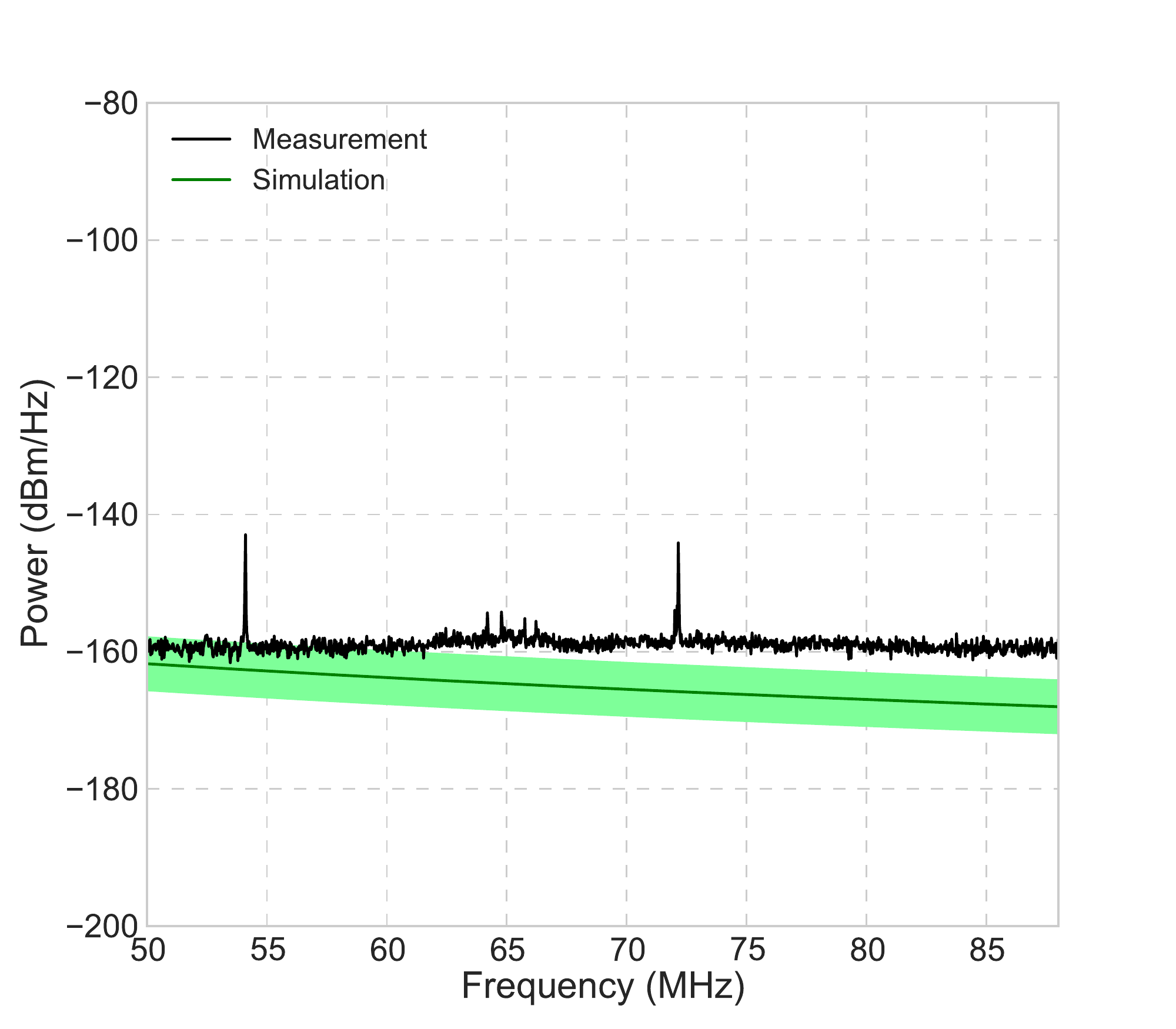}}}
\caption{Measured RS1 noise floor, with estimated galactic noise overlaid in green. Data taken January,
2015.
\label{fig:RS1_noisefloor}}
\end{figure}

\begin{figure}[!h]
\centerline{\mbox{\includegraphics[width=0.48\textwidth]{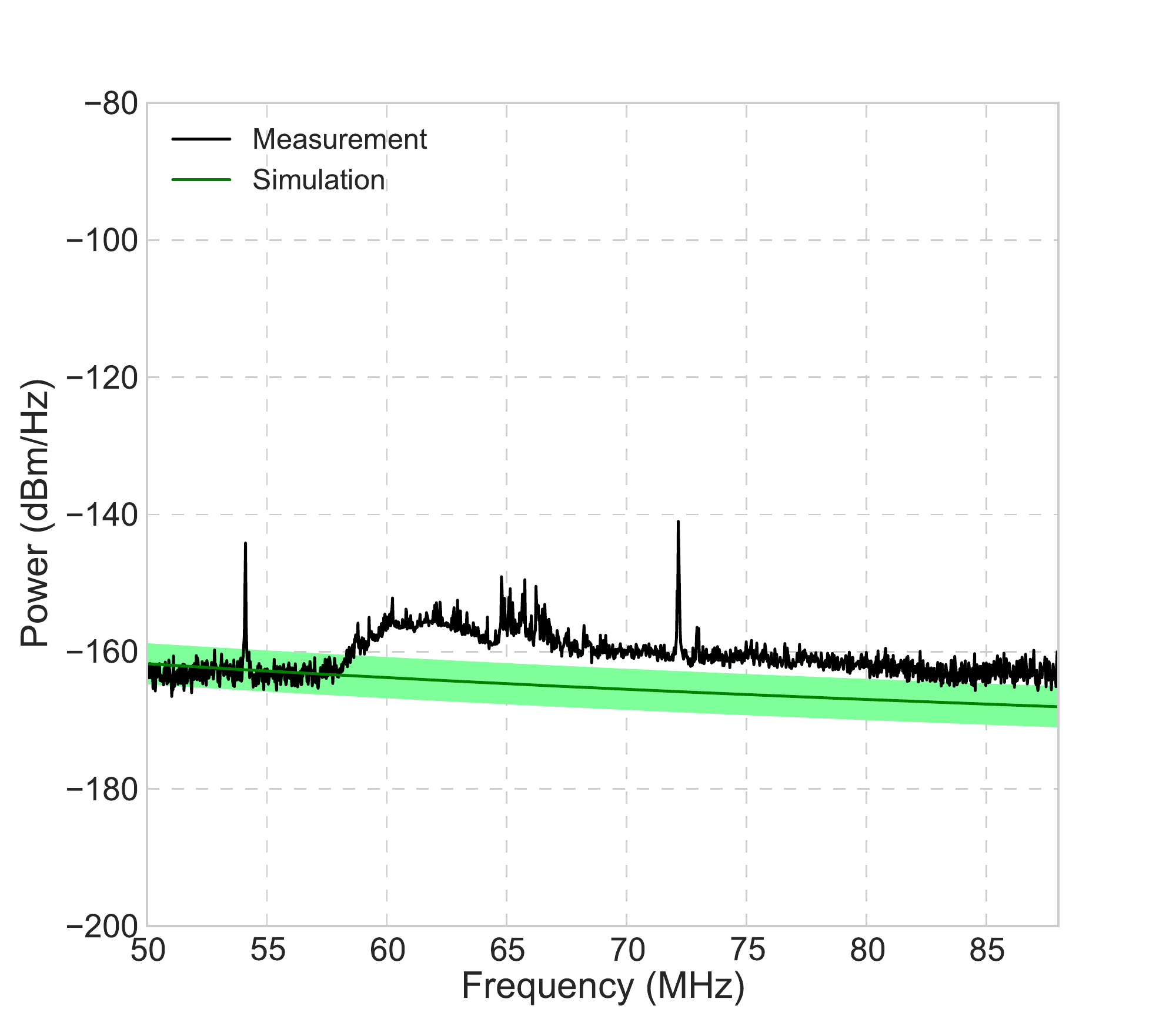}}}
\caption{Measured RS2 noise floor, with estimated galactic noise overlaid in green. Data taken January,
2015.
\label{fig:RS2_noisefloor}}
\end{figure}

Currently, trigger rates for both stations are well below 1 Hz, and therefore allow operation at nearly 100\% livetimes. Since upgrading RS-1 and installation of RS-2, our analysis has thus far focused on a verification that the ambient noise floor is approximately commensurate with expectations from a combination of thermal plus galactic noise. Figures \ref{fig:RS1_noisefloor} and \ref{fig:RS2_noisefloor} show the averaged 
corrected VPol power for each of the two remote stations based exclusively on forced triggers, respectively. Despite considerable filtering, including the custom RLC filter that provides a 24 dB deep notch over a 1 MHz bandwidth, the carrier at 54.1 MHz is still observable above the ambient background. We also observe a local 72.2 MHz source of unknown origin. Aside from those continuous wave (CW) lines, the ambient background is within the few dB absolute systematic error (for each station independently) of our expectation for the irreducible background in this frequency band.

\begin{figure}[!h]
\centerline{\mbox{\includegraphics[width=0.48\textwidth]{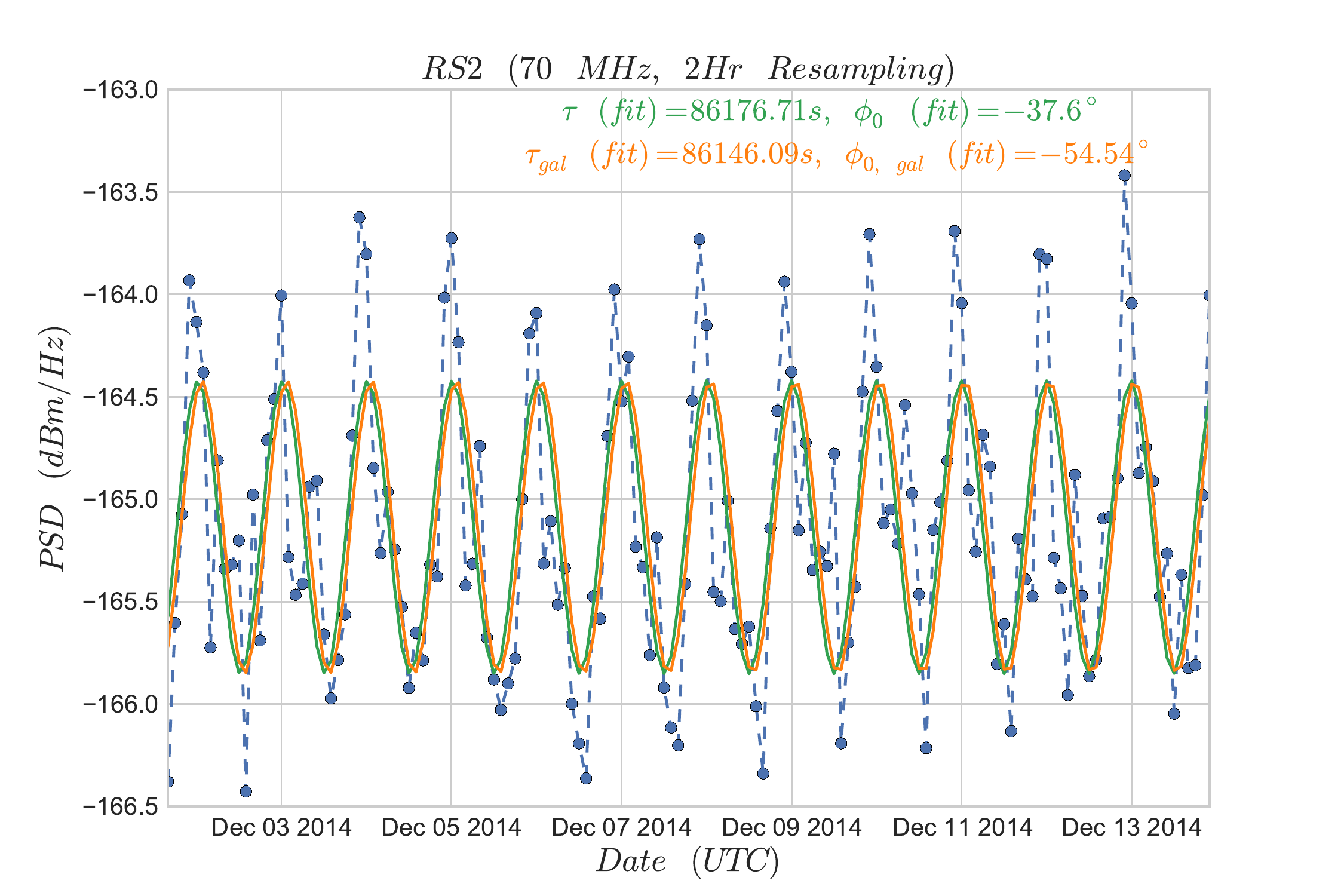}}}
\caption{Variation in measured ambient environmental power (taken in a 5 MHz band 
centered on 70 MHz, superimposed with the movement of the galactic center in the sky. 
The former has been fit to a sine wave, giving agreement with the expected phase of the galactic center to within 5\%.
\label{fig:MilkWeg}}
\end{figure}

The RS environment should be sufficiently radio-quiet that the sidereal variation in
the environmental power, due to the motion of the galactic center across the sky, should
be observable. Figure \ref{fig:MilkWeg} shows the measured environmental power,
for a 5 MHz band centered at 70 MHz, as a function of time for the first 12 days of RS-2 operation. Overlaid is
a fit to that variation, as well as the known azimuthal position in the sky of the galactic
center relative to boresight of the front-end log-periodic dipole
antennas. As expected, the two curves are very close to each other in absolute phase.

\section{Summary and Outlook}
As of mid-November, 2014, the TARA transmitter has been rotated from its initial HPol configuration into a VPol configuration, designed to better match the preferential polarization from directly vertically descending showers. Data will be accumulated for approximately one year with the current TARA configuration, consisting of a National Instruments Flex-RIO-based DAQ located at the main TARA data-taking building in LongRidge, UT, plus the two remote stations described herein. As mentioned herein, the radar cross-section is directly related to the size of the reflective plasma along the shower axis (``shower core''); that plasma density is predicted by sophisticated air shower simulation packages such as CORSIKA. If the radar cross-section is equivalent to that predicted by CORSIKA simulations for a typical 10 EeV UHECR-induced extensive air shower, these stations should observe tens of events per month at the -10 dB trigger threshold achieved in the lab. Conversely, the lack of definitive UHECR observations via radar reflections will result in constraints on the effective air shower radar cross-section, or, alternately, the lifetime of the putative shower core reflective ionization plasma. In such a case, mechanisms such as collisional damping of the coherent plasma response due to interactions with local gas molecules, or even loss of charge in the core plasma due to recombination of plasma charge with gas molecules (likely oxygen, which has a higher electron affinity than nitrogen) may render an unfavorable outlook for the continued development of this approach in-air. Nevertheless, the technique may still offer promise for application to showers developing in solid media such as ice, as has been suggested elsewhere\cite{IceRF}.

\section{Acknowledgments}
This work is supported by the U.S. National Science Foundation Grant nos. NSF/PHY-0969865 and NSF/MRI-1126353, 
by the Megagrant 2013 program of Russia, via agreement 14.А12.31.0006 from 24.06.2013,
by the Vice President for Research of the University of Utah, and by the W.M. Keck Foundation. 
D.B., A.N., A.S. and M.S. acknowledge support from
National Research Nuclear University MEPhI (Moscow Engineering Physics Institute).
L.B. acknowledges the support from NSF/REU-1263394. We would also like to acknowledge the generous donation of analog television transmitter equipment by Salt Lake City KUTV Channel 2 and ABC Channel 4, and the cooperation of Telescope Array collaboration.


\section{References}
\begin{thebibliography}{99}
\bibitem{GZK}
  K.~Greisen,
  ``End To The Cosmic Ray Spectrum?,
  Phys.\ Rev.\ Lett.\  {\bf 16}, 748 (1966);
  G.~T.~Zatsepin and V.~A.~Kuzmin,
  ``Upper Limit Of The Spectrum Of Cosmic Rays,''
  JETP Lett.\  {\bf 4}, 78 (1966)
  [Pisma Zh.\ Eksp.\ Teor.\ Fiz.\  {\bf 4}, 114 (1966)].

\bibitem{PAO}E. Roulet, ``Recent Results from the Pierre Auger Observatory'', Proc. X SILAFAE Medellin-2014 (2015), arXiv:1503.09173
\bibitem{TA}M. Fukushima, ``TA Recent Results and Prospects'', Proc. of Intl. Symp. on Very High Energy
Cosmic Rays (ISVHECRI2014), CERN, (2014), arXiv:1503.06961 
\bibitem{AERA}J. Horendel, ``Radio detection of air showers with LOFAR and AERA'', Proc. of Ultra-High Energy Cosmic Ray Conference, Springdale, UT (2014).
\bibitem{Helio}H. Takai, ``Forward Scattering Radar for Ultra-High Energy Cosmic Rays'', Proc. of 32nd Intl. Cosmic Ray Conference (2011).
\bibitem{Isaac14}R. Abbasi {\it et al}, ``Telescope Array Radar (TARA) Observatory for Ultra-High Energy Cosmic Rays'', Nucl. Instr. and Meth. in Physics Research {\bf A}, 767, 322 (2014).
\bibitem{GH}T.K. Gaisser and A.M. Hillas, ``Reliability of the method of constant intensity cuts for reconstructing the average development of vertical showers''. Proc. of 15th Int. Cosmic Ray Conf., 13–26 (1977).
\bibitem{Gorham}P.W. Gorham {\it et al}, ``Observations of Microwave Continuum Emission from Air Shower Plasmas'', Phys. Rev. {\bf D78} 032007 (2008). 
\bibitem{SamThesis}S. Kunwar, ``Design And Development Of An Autonomous Radar Receiver For The Detection Of Ultra High Energy Cosmic Rays'', Ph.D. Thesis, University of Kansas (2015).
\bibitem{AD80066}http://www.analog.com/en/products/digital-to-analog-converters/da-converters/ad80066.html
\bibitem{AD9634}http://www.analog.com/ru/products/analog-to-digital-converters/ad-converters/ad9634.html
\bibitem{Nexsys3}www.digilentinc.com/nexys3
\bibitem{XilinxUG381}http://www.xilinx.com/support/documentation/user\_guides/ug381.pdf
\bibitem{RaspPi}http://respberrypi.ru
\bibitem{WiringPi}http://wiringpi.com
\bibitem{ilotus}http://www.ilotus.com.sg
\bibitem{IceRF}K. D. deVries {\it et al.}, ``Coherent Transition Radiation in Askaryan radio detectors'', http://xxx.lanl.gov/abs/1503.02808, submitted to Astropart. Phys. (2015)
\end{thebibliography}
\end{document}